\documentclass{aastex63}

\newcommand{\mjup} {\ensuremath{M_{\rm Jup}}}
\newcommand{\Msun}{\ensuremath{M_{\odot}}}
\newcommand{\um}{\ensuremath{\mu{\rm m}}}
\newcommand{\HST}{{\sl HST}}

\usepackage{comment}
\usepackage[figure,figure*]{hypcap}
\usepackage{placeins}

\received{April 16, 2023}
\accepted{October 1, 2023}
\submitjournal{ApJ}

\shorttitle{The substellar population of NGC~2024}
\shortauthors{Robberto et al.}
\graphicspath{{./}{figures/}}

\begin{document}

\title{A HST study of the sub-stellar population  of NGC~2024}

\correspondingauthor{Massimo Robberto}
\email{robberto@stsci.edu}

\author[0000-0002-0786-7307]{Massimo Robberto}
\affiliation{Space Telescope Science Institute, 3700 San Martin Dr, Baltimore, MD 21218, USA}
\altaffiliation{
\begin{minipage}{28em}
Based on observations made with the NASA/ESA {\it Hubble
Space Telescope}, obtained at the Space Telescope Science Institute, which
is operated by the Association of Universities for Research in Astronomy, 
Inc., under NASA contract NAS 5-26555.  These observations are associated
with program GO-13826.
\end{minipage}
}
\affiliation{Johns Hopkins University, 3400 N. Charles Street, Baltimore, MD 21218, USA}

\author[0000-0002-5581-2896]{Mario Gennaro}
\affiliation{Space Telescope Science Institute, 3700 San Martin Dr, Baltimore, MD 21218, USA}
\affiliation{Johns Hopkins University, 3400 N. Charles Street, Baltimore, MD 21218, USA}

\author{Nicola Da Rio}
\affiliation{Dept. of Astronomy, University of Virginia, Charlottesville, Virginia, USA}

\author[0000-0002-1652-420X]{Giovanni Maria Strampelli}
\affiliation{Johns Hopkins University, 3400 N. Charles Street, Baltimore, MD 21218, USA}
\affiliation{Space Telescope Science Institute, 3700 San Martin Dr, Baltimore, MD 21218, USA}
\affiliation{Department of Astrophysics, University of La Laguna, Av. Astrofísico Francisco Sánchez, 38200 San Cristóbal de La Laguna, Tenerife, Canary Islands, Spain}


\author{Leonardo Ubeda}
\affiliation{Space Telescope Science Institute, 3700 San Martin Dr, Baltimore, MD 21218, USA}

\author[0000-0003-2954-7643]{Elena Sabbi}
\affiliation{Space Telescope Science Institute, 3700 San Martin Dr, Baltimore, MD 21218, USA}

\author[0000-0002-1652-420X]{Dana Koeppe}
\affiliation{Johns Hopkins University, 3400 N. Charles Street, Baltimore, MD 21218, USA}

\author[0000-0002-3389-9142]{Jonathan C. Tan}
\affil{Dept. of Space, Earth \& Environment, Chalmers University of Technology, Gothenburg, Sweden}
\affiliation{Dept. of Astronomy, University of Virginia, Charlottesville, Virginia, USA}

\author[0000-0002-0322-8161]{David R. Soderblom}
\affiliation{Space Telescope Science Institute, 3700 San Martin Dr, Baltimore, MD 21218, USA}

\begin{abstract}
We performed a HST/WFC3-IR imaging survey of the young stellar cluster NGC 2024 in three filters probing the 1.4~$\mu$m H$_2$O absorption feature, characteristic of the population of low mass and sub-stellar mass objects down to a few Jupyter masses. We detect 812 point sources, 550 of them in all 3 filters with signal to noise { greater than 5}. Using a distance-independent two-color diagram { we determine extinction values as high as $A_V\simeq 40$. We also find that the change of effective wavelengths in our filters results in higher $A_V$ values as the reddening increases}. Reconstructing a dereddened color-magnitude diagram we derive a luminosity histogram for both the full sample of { candidate} cluster members and for an extinction-limited sub-sample containing the 50\% of sources with $A_V\lesssim 15$. { Assuming a standard extinction law like \cite{Cardelli_1989ApJ...345..245C} with a nominal $R_V$=3.1 we produce a luminosity function in good agreement with the one resulting from a Salpeter-like Initial Mass Function for a 1~Myr isochrone. There is some evidence of an excess of luminous stars in the most embedded region}. We posit that the correlation {  may be due to those sources being younger}, and therefore overluminous than the more evolved and less extinct cluster’s stars. We compare our classification scheme based on the depth of the 1.4~$\mu$m photometric feature with the results from the spectroscopic survey of Levine et al. (2006), and we report a few peculiar sources and morphological features typical of the rich phenomenology commonly encountered in young star-forming regions.
\end{abstract}


\section{Introduction} \label{sec:intro}
Observations of the youngest stellar clusters in the solar vicinity provide unique insights on the star formation phenomenon \citep{LadaLada2003ARA&A..41...57L}. Stellar systems with ages shorter than their dynamical time-scales preserve the original characteristics imprinted by the star formation process, allowing to directly compare the basic physical and dynamical parameters of their members with those of { pre-stellar cores} in Giant Molecular Clouds, and with theoretical models \citep{Krumholz+2019ARA&A..57..227K}.  On the other hand, rich young clusters may be strongly affected by feedback from { their} massive stars.
{Through photoevaporation { and stellar winds}, the copious amount of { radiative and mechanical energy } released by massive stars plays an accumulative role in the star and planet formation process \citep{1999ApJ...515..669S,  Dib+2010, 2020MNRAS.491..903W} as well as in the cluster dynamics through the dispersal of diffuse material and the resulting decrease of gravitational binding energy \citep{10.1093/mnras/sty035} { with the loss of more than 50\% of stars \citep{Brinkmann2017}. At the same time, the high stellar density at the cluster center may cause close encounters with explosive ejections of runaway stars \citep{Rivera-Ortiz+2021} as well as protoplanetary disk fragmentation, again combined with ejection \citep{Whitworth+2007prpl.conf..459W, Bonnell2008,Stamatellos+2009}. { The majority of low mass stars are especially sensitive to these effects, and measuring their global properties, with their variations both within and between different clusters, may clarify the frequency and relevance of these complex physical processes}.

A key parameter is the the low-mass IMF, often parameterized using the canonical Chabrier or Kroupa forms established for the Milky Way disk. In the 
substellar regime, howerver, the IMF is  difficult to characterize due to a multitude of factors \citep{Offner+2013, Hopkins2018}. This is reflected in the original broken power law of \cite{Kroupa2001}  which has an $\alpha$ exponent with large uncertainty, i.e. $\alpha=0.3\pm0.7$ in the mass range $0.01\le m/\Msun\le0.08$. While studies of different clusters in the solar vicinity, e.g. the Orion Nebula Cluster  \citep{Gennaro&Robberto2020ApJ...896...80G},  { NGC~2244}, \citep{Almendros-Abad+2023arXiv230507158A}, Chamaeleon I \citep{Luhman2007}, $\lambda$ Ori \citep{Bayo+2011}, $\sigma$ Ori \citep{Pena-Ramirez+2012, Damian+2023_arXiv230317424D}, $\rho$ Ophiuchi \citep{Alves_De_Oliveira+2012}, Upper Scorpius \citep{Lodieu2013}, NGC~1333 and IC~348 \citep{Scholz+2013ApJ...775..138S}, Lupus 3 \citep{Muzic+2015ApJ...810..159M},  25Ori \citep{Downes+2014MNRAS.444.1793D, Suarez+2019MNRAS.486.1718S} report IMF generally consistent with the canonical laws, a number of studies \citep{Weidner+2013, Dib2014, Dib+2017} provide statistically-significant evidence of variations of IMF between clusters. A spread of IMF parameters may actually explain the integrated IMF of galaxies of different metallicity \citep{Dib2022}.}

Among the closest young clusters, the one associated to the 
``Flame Nebula'', NGC~2024, plays a special role, being often regarded as a young, down-sized version of the more extensively studied Orion Nebula Cluster (ONC). { Both} are located at nearly the same $\sim400$~pc distance and in the same complex of molecular clouds, NGC~2024 { representing} the richest star-forming region in the Orion-B molecular cloud \citep{Meyer+2008hsf1.book..662M}, while the ONC is the richest young cluster in the Orion-A molecular cloud.
The age of NGC~2024, $\sim 0.5\,{\rm Myr}$ as determined by early near-infrared photometric studies \citep[e.g.,][]{1996ApJ...473..294C, 1996PhDT.........3M, 2000AJ....120.1396H} is smaller than that of the ONC ($\sim 1-3\,{\rm Myr}$), consistent with the fact that NGC~2024 appears more heavily obscured. Its morphology at visible wavelengths is dominated by a North-South dark lane of high opacity \citep[$A_V\simeq20$ mag,][]{2003ApJ...598..375S}, { the densest part of a foreground dust layer} causing extreme differential extinction across the field \citep{Barnes+1989_ApJ...342..883B}. { Performing} photometry at near-IR wavelengths, 
{ \cite{1996PhDT.........3M} }
found that the majority of cluster members,  $\simeq 70\%$, harbor accreting disks. 
{ Using the NICMOS-3 camera onboard the Hubble Space Telescope (\HST)~\cite{Liu+2003} imaged the inner 4.27 square arcminutes of NGC ~2024 in the F110W and F160W filters. Their photometry of 79 sources indicates a ratio of intermediate to low mass objects consistent with the field IMF.}
Following the photometric observations, \cite{Levine+2006} spectroscopically probed the low-mass/substellar candidates, finding that out of 70 spectra, $\sim 15$ can be interpreted as substellar objects. Penetrating the dust lane at longer wavelengths, Spitzer/IRAC images unveiled that the young stellar cluster reaches a stellar density $n_*\simeq 2000$~pc$^{-3}$ 
\citep{2005IAUS..227..383M}. Rich, young clusters are typically dominated by a few massive stars. In the case of NGC 2024, the main ionizing source is currently considered to be IRS 2b, a highly reddened late-O type star \citep[][reported an $\sim$O8 spectral type]{Bik+2003A&A...404..249B} to the East of the dark lane and slightly off-center with respect to the bulk of the general young
stellar population. 
More recently, \cite{2020A&A...640A..27V} used ALMA to survey at 1.3~mm in the central $2.9^\prime\times2.9^\prime$ region detecting 179 disks. They find that disks at the smallest projected distance from IRS 2b appear younger and more massive than those to the west of the dark lane, a region of lower extinctions dominated by a less massive B0.V star, IRS 1. 
Using archival HST images, \cite{Haworth+2021MNRAS.501.3502H} showed that photoevaporated disks (proplyds) can be found in both regions, with ionized cusps pointing to both IRS 1 and IRS 2b, indicating that both sources are primary actors in the external photoevaporation of disks in NGC~2024.

In order to improve the census of the NGC~2024 cluster in the substellar mass regime, we have performed a near-IR imaging survey adopting a { multi-band photometry} technique that allows estimating the temperature of sources with $T_{\rm eff}\lesssim3200$~K. Compared to the more accurate infrared spectroscopy, multi-band photometry has two main advantages: 1\textit{}) observing efficiency, as hundreds of sources can be simultaneously observed with high signal-to-noise, and 2) lack of bias, as the entire cluster can be sampled down to a certain (mass-dependent) extinction level, without pre-selection of the targets. 

The primary spectral features probing the temperature range typical of low-mass stars and brown dwarfs are arguably the H$_2$O and CH$_4$ near-infrared bands. These bands are hardly measurable from the ground because of the strength and variability of telluric absorption, but are fully accessible with space-based instruments like the WFC3/IR on the Hubble Space Telescope. { In a previous study of the ONC, \cite{Robberto-ONC-2020ApJ...896...79R} used} the WFC3/IR F130N and F139M photometric bands { to build a spectral index} tracing the depth of the H$_2$O absorption feature at $\lambda\simeq1.4\mu$m. { In the case of the ONC, this technique  allows to} efficiently disentangle the population of low-mass stars from the contamination of Galactic and extragalactic objects, which is significant at the faint magnitude levels of the young planetary mass objects. 

Here, we present the results obtained by performing a similar survey in the NGC~2024 region. 
For this study, a third filter close in wavelength to our other two filters, the “wide Y” F105W bandpass, has been added to produce a distance-independent two-color diagram useful to assess the substantial reddening toward each source while minimizing the risk of contamination from { thermal} IR excess due to circumstellar disks.

This paper is organized as follows: in Section~\ref{sec:observations} we present our observing strategy and data reduction methodology. In Section~\ref{sec:results} we present our results, i.e., the main photometric catalog and the Color-Magnitude Diagram (CMD), the extinction estimated from the Color-Color diagram and the Luminosity Function from the dereddened version of a CMD. In Section~\ref{sec:Levine}
we compare the quantities extracted from our 1.4~$\mu$m index with those derived using spectroscopy by \cite{Levine+2006}. Finally, in Section~\ref{sec:conclusion} we summarize our findings, whereas in the Appendix we introduce a few noticeable objects and morphological features revealed by our visual inspection of the images.

\section{Observations and Data Processing} \label{sec:observations}
The observations presented here are part of the HST GO-15334 program (Principal Investigator N. Da Rio). They have been executed with the Wide Field Camera 3 onboard the Hubble Space Telescope between October 2018 and April 2020. Twenty orbits were allocated to produce a mosaic of $5\times4$ WFC3/IR pointings, as shown in Figure~\ref{fig:Footprint}. To maximize areal coverage { and maintain uniformity of exposure time across the field}, the overlap between individual tiles was kept to a minimum. However, the original plan of building a mosaic without gaps was hampered by the difficulty of finding suitable guide stars at the { optimal} telescope pointings and orientations. Iterating with the HST program coordinators at the Space Telescope Science Institute, we were finally able to craft a configuration that leaves only a small fraction of the field uncovered. Overall, the survey extends over 84.57 square arcminutes centered on RA=05:41:42.945 and DEC=$-$1:54:18.58 (J2000). For comparison, this is about 17\% of the area covered by the HST-GO13826 program on the ONC. 

\begin{figure*}
\centering
\includegraphics[width=0.75\textwidth]{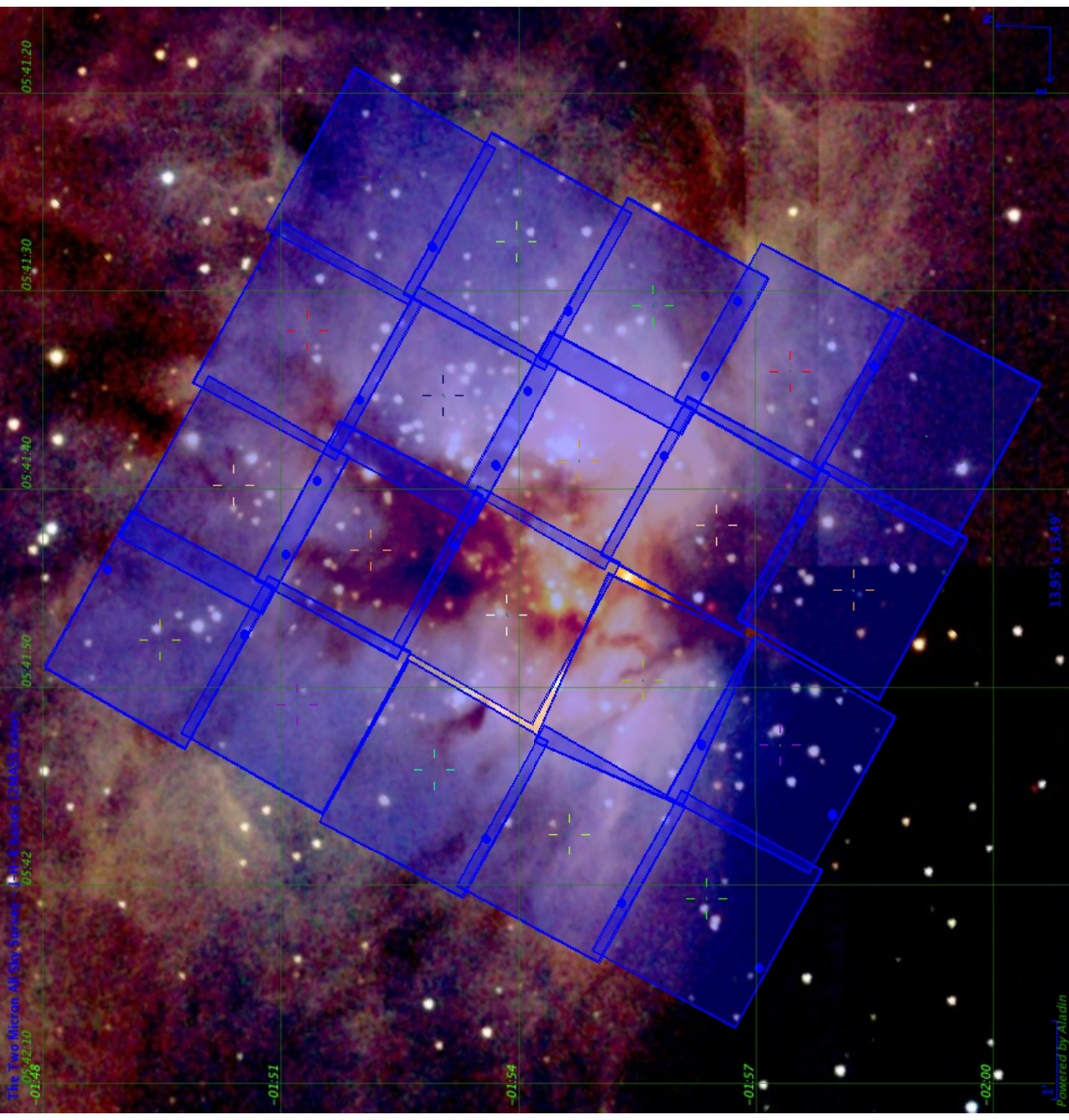}
\caption{Footprint of the NGC~2024 dataset presented in this paper, overlaid on the 2MASS image of the field.}
\label{fig:Footprint}
\end{figure*}

The filters used were F130N, F139M and F105W. The exposure times were different for each filter and were determined considering the different widths of the three passbands in order to achieve similar signal-to-noise in all filters. For the narrow-band F130N filter, the longer exposure time was split in 2 halves to limit the influence of cosmic rays. At each pointing, three slightly dithered exposures were taken for each filter to mitigate against bad pixels and cosmic rays. The readout parameters are presented in Table~\ref{tab:Readout}.

\begin{deluxetable}{cccccccc}
 \tablecaption{WFC3 detector readout parameters\label{tab:Readout}}
 \tablehead{
\colhead{Filter} 
        & \colhead{Readout Pattern}
                    & \colhead{N$_{\rm samp}$} 
                        & \colhead{Exp. time/pointing} 
                                            &\colhead{Exp.time/visit}
 }
\startdata
F130N   & SPARS-50 & 7  & $2\times303$ s    & 1,817 s\\
F139M   & SPARS-25 & 6  & 128 s             & 385.8 s\\
F105W   & SPARS-10 & 5  & 43 s              &  128.8 s\\
\enddata
\vspace{-0.5cm}
\end{deluxetable}

\par\vfill\eject
The images were registered against the GAIA-DR1\citep{GAIA-DR1b-2016A&A...595A...1G, Gaia-DR1a-2016A&A...595A...2G}\footnote{GAIA-DR1 was the version available when data were processed. Since the differences between DR1 and DR2 are negligible for our goals, we have maintained our original astrometric solution throughout the analysis work.} catalog and drizzled to the nominal WFC3/IR pixel scale of 0.12825 arcsecond/pixel.
Figure~\ref{fig:mosaic_color} shows the final mosaic obtained by combining the three filters as RGB color composite image.
\begin{figure*}
\centering
\includegraphics[width=0.95\textwidth]{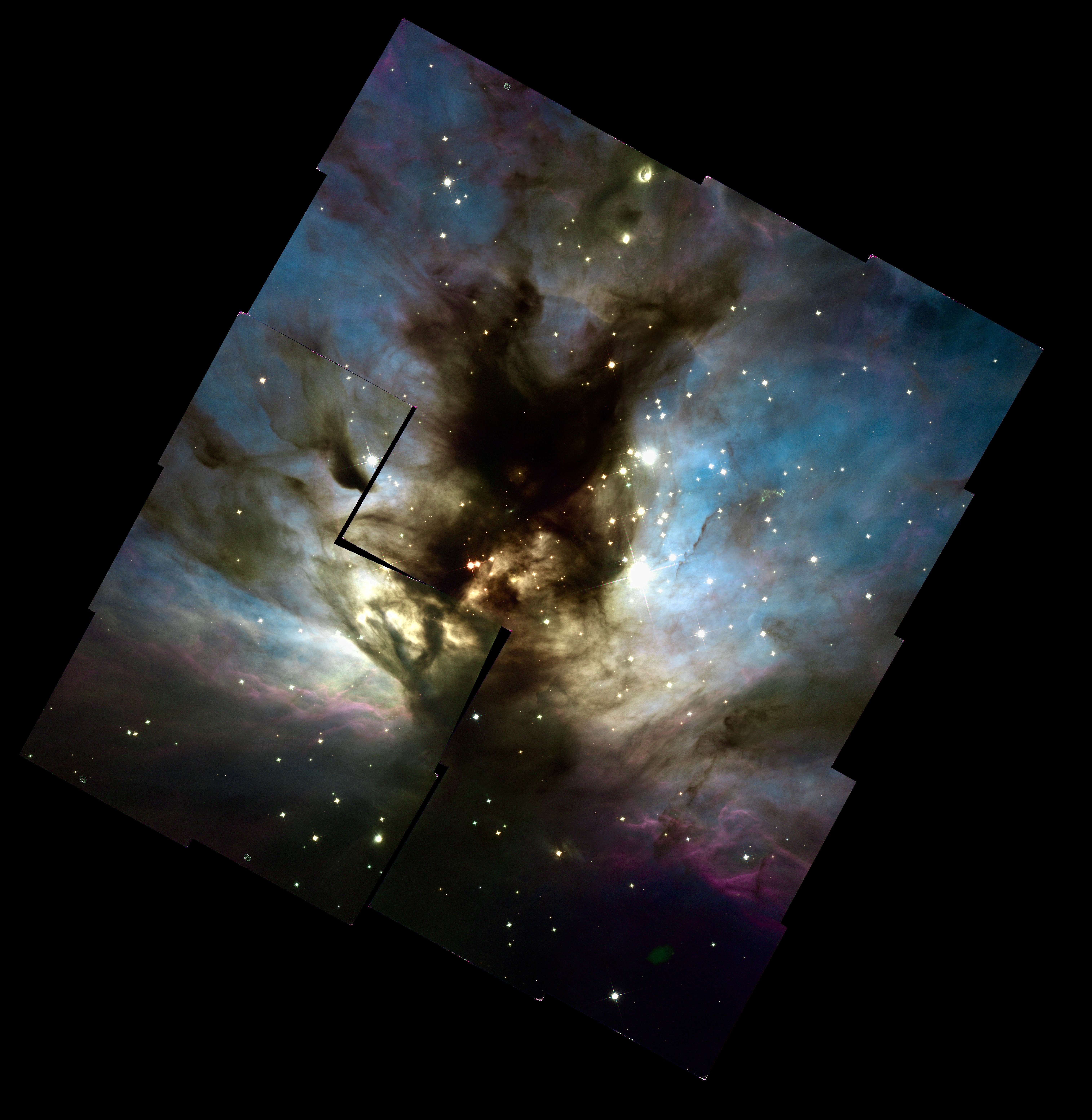}
\caption{Color-composite mosaic obtained by combining in the RGB planes the F139M, F130N  and F105W drizzled images, in the order.}
\label{fig:mosaic_color}
\end{figure*}

\par\vfill\eject
{ An initial source catalog
was obtained using the Dolphot package \citep{DOLPHOT_2000, DOLPHOT_2016} on the 20 dithered exposures of each visit. 
Visual inspection of the final drizzled images was necessary to eliminate residual artifacts due to persistence and confirm a few low signal-to-noise objects.} 
We present the results obtained performing aperture photometry on the final drizzled images, with an extraction aperture of 0.4 arcsec and the most recent (year 2020) release of the WFC3-IR zero points \citep{WFC3_ZeroPoints}. Since the new zero points are provided for infinite aperture, we adopted the aperture corrections given by the previous 2012 calibration. As a result, the magnitudes determined using a 0.4 arcsec aperture radius on the drizzled images were corrected to Vega magnitudes using the following zero points: 
ZP(130N)=21.797,  
ZP(F139M)=23.175, 
and
ZP(F105W)=25.432. 

Figure~\ref{fig:dmag_vs_mag} shows the distribution of  photometric uncertainty, $dm$, vs. magnitude, $m$, in the Vega system { for  the sources detected with $dm< 0.3$~mag (SNR$\gtrsim 3$). These are 618 source in the F105W,  761 in the F130N and 800 in the F139M filters.}
%
\begin{figure*}
\centering
\includegraphics[width=0.95\textwidth]{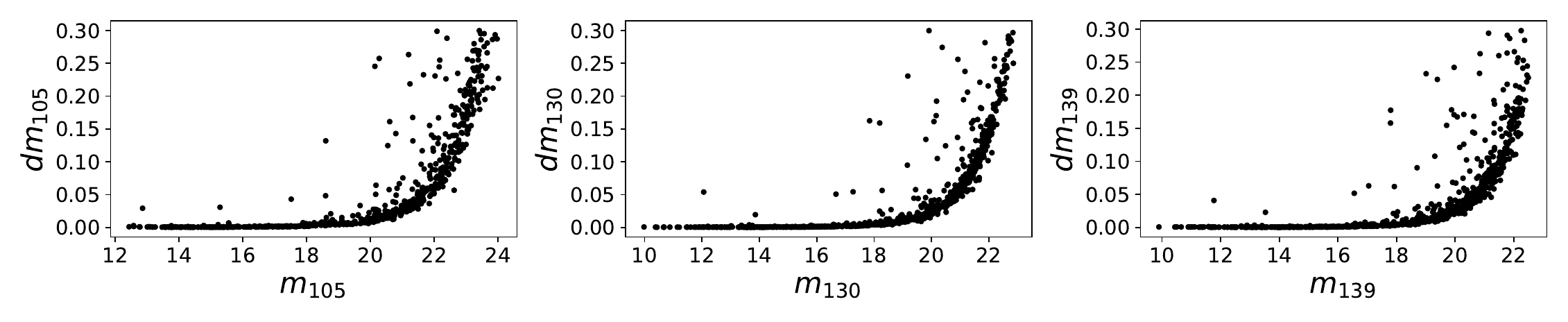}
\caption{Distribution of the photometric uncertainty vs. magnitude for the three filters, Vega system.}
\label{fig:dmag_vs_mag}
\end{figure*}

{ Our sensitivity limits are not uniform across the field. To assess their spatial variation we randomly injected 90,000 sources with magnitudes in the range 21-25,  retrieved their flux using aperture photometry as for the real sources, and estimated the number of sources measured with SNR$>3$ at different magnitude thresholds. 
Figure~\ref{fig:area_fraction} shows how the fraction of recovered sources drops moving to fainter magnitudes, reaching 50\% at approximately $m_{105W} =23.5$, $m_{130N} =22.4$, and $m_{139M} =21.6$ on average across the field, but with 
significant spatial variations.

\begin{figure*}
\centering
\includegraphics[width=0.65\textwidth]{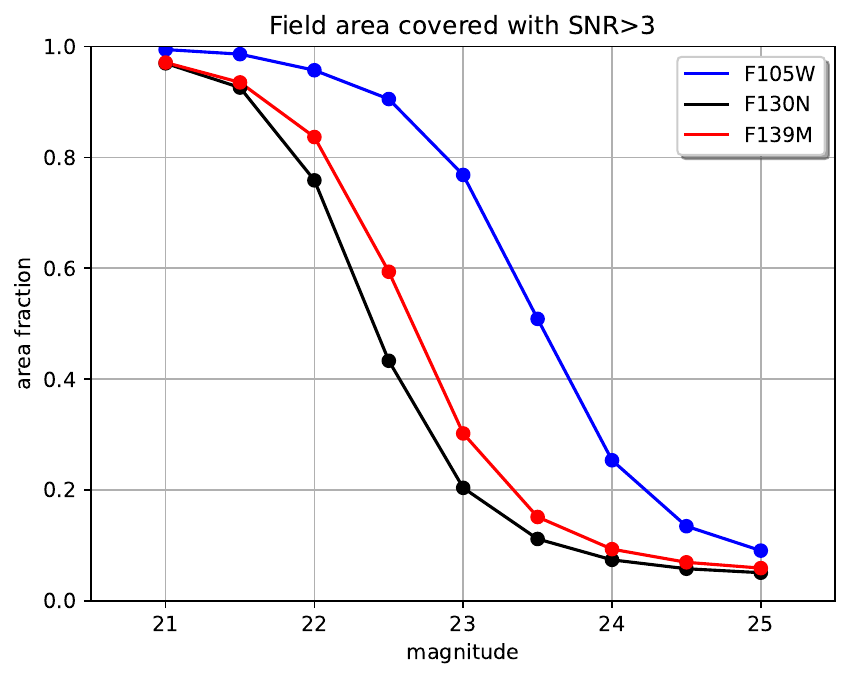}
\caption{ Fraction of the survey area covered with $SNR > 3$ vs. magnitude in the three passbands. }
\label{fig:area_fraction}
\end{figure*}
The map in Figure~\ref{fig:3magnitudeplots}, relative to the F105W filter, shows how the sensitivity at the 3$\sigma$ level decreases by a couple of magnitudes in correspondence of the brightest regions due to the increased background. Similar maps are obtained for the other two filters. In practice, the darkest regions, those with the lowest source density due to high extinction, are also those where our detection threshold increases, and vice-versa for the brightest regions. Therefore, the completeness of our source catalog, intended as the number of sources detected vs the number of cluster members, is predominantly determined by extinction to the cluster and cannot be recovered using a sensitivity map, unlike when completeness is determined e.g. by source confusion.}

{ At the bright end, all sources are accounted for in our catalog with the exception of source IRS~5 of \cite{Barnes+1989_ApJ...342..883B} as it falls right in a gap between our tiles. This is one of the brightest near-infrared sources in the field  with 2MASS magnitudes J=10.35, H=8.53 and K=7.54 \citep{2MASS_2003yCat.2246....0C}. }


\begin{figure}[htp]

\centering
\includegraphics[width=.80\textwidth]{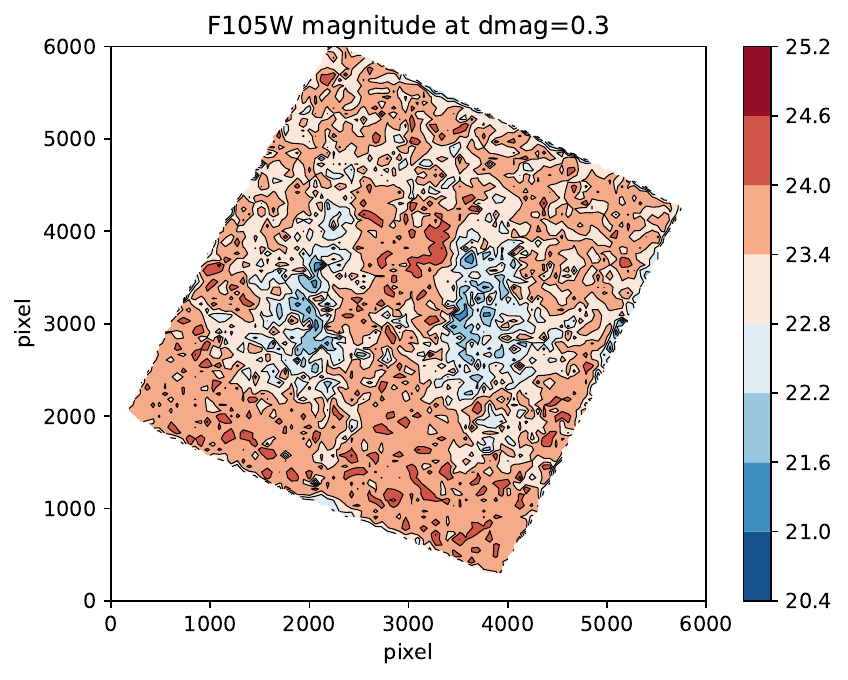}\hfill
\caption{Spatial distribution of the limit magnitude in the F105W filter.}
\label{fig:3magnitudeplots}
\end{figure}


The full photometric catalog, containing { 812} entries
is presented in Table~\ref{tab:Photometry}.  From left to right, the columns show our entry number, the Right Ascension and Declination at the J2000.0 epoch, and the Vega-system magnitudes and associated uncertainties in the F105W, F130N and F139M filters. 
In the following sections  we shall concentrate on the 550 
sources with magnitude uncertainties $<0.2$~mag in all three filters.

\begin{deluxetable}{ccccccccc}
\tablecaption{WFC3-IR Photometry from  the HST GO-15334 program\label{tab:Photometry}}
\tablehead{
\colhead{Entry nr.}&
\colhead{RA(J2000)}&
\colhead{Dec(J2000)}&
\colhead{m\_105}&
\colhead{dm\_105}&
\colhead{m\_130}&
\colhead{dm\_130}&
\colhead{m\_139}&
\colhead{dm\_139}
}
\startdata
1   & 85.396685& -2.008753  & 22.640    & 0.156 & 21.507& 0.102 & 21.347&   0.094   \\
2   & 85.396449& -2.008472  & 23.373    & 0.230 & 21.383& 0.076 & 21.285&   0.073   \\
3   & 85.398049& -2.007868  & 19.287    & 0.007 & 17.755& 0.004 & 17.357&   0.002  \\
... & ...      & ...       & ...        & ...   & ...   & ...   & ...   &   ...     \\
\enddata
\tablecomments{The full table is available for download in the electronic version of the paper.}
\end{deluxetable}


\par\vfill\eject

\section{Results}\label{sec:results}
\subsection{Color-Magnitude Diagrams}

Combining the three filter pairs, one can produce the CMDs presented in 
Figure~\ref{fig:3CMDs}. On each CMD we overplot the unreddened 1~Myr  isochrones { at 400~pc} from the BT-Settl family of models { (blue dashed line)}
with four representative points corresponding to a 0.4, 0.072, 0.015, and 0.002~\Msun\, star, top to bottom. Assuming no extinction, the highest value roughly corresponds to our saturation limits, whereas the lowest value corresponds to our detection limits and the intermediate values are representative of the hydrogen and deuterium burning limits. 

{ The first of the three diagrams, the F130N-F139M CMD, presents on the horizonal axis our H$_2$O index and was derived  also for the ONC by \cite{Robberto-ONC-2020ApJ...896...79R}. They found that the BT-Settl isochrone calculated down to 0.5~\mjup\, 
predicts highly negative (blue) H$_2$O color as one approaches the lowest masses. Our new diagram for NGC~2024 is consistent with those findings, in the sense that there are no sources with color F130N-F139M bluer than -0.5. 
To match the ONC observations, \cite{Robberto-ONC-2020ApJ...896...79R} applied a correction  to the magnitude of the BT-Settl isochrone in the F130N filters, and we shall adopt here the same recipe. The discrepancy between theory and observations} might be attributed to the incomplete line lists and opacity tables used to predict the water absorption feature for very low effective temperature and surface gravity objects. Historically, the shape of the spectra of very young and very low mass objects in the 1.4$\mu$m region has always been difficult to probe from the ground due to telluric water vapor absorption. The paucity of observational benchmarks reflects on the accuracy of theoretical models at these wavelengths, a situation that is certainly going to change once new JWST observations become available.  

{ The slope of the reddening vectors, also shown in the figure,  have been determined using Synphot \citep{SYNPHOT_2018} to calculate  the change of the Vega spectrum in our passbands adopting the Milky Way $R_V=3.1$ reddening law by \cite{Cardelli_1989ApJ...345..245C}}. 
We estimate that an extinction $A_V=5$~mag in the standard Johnson V band-pass corresponds to $A_{105W}=1.88$, $A_{130N}=1.31$ and $A_{139M}=1.18$ mag, and the lines have been drawn according to these values.  

\begin{figure*}
\centering
\includegraphics[width=0.95\textwidth]{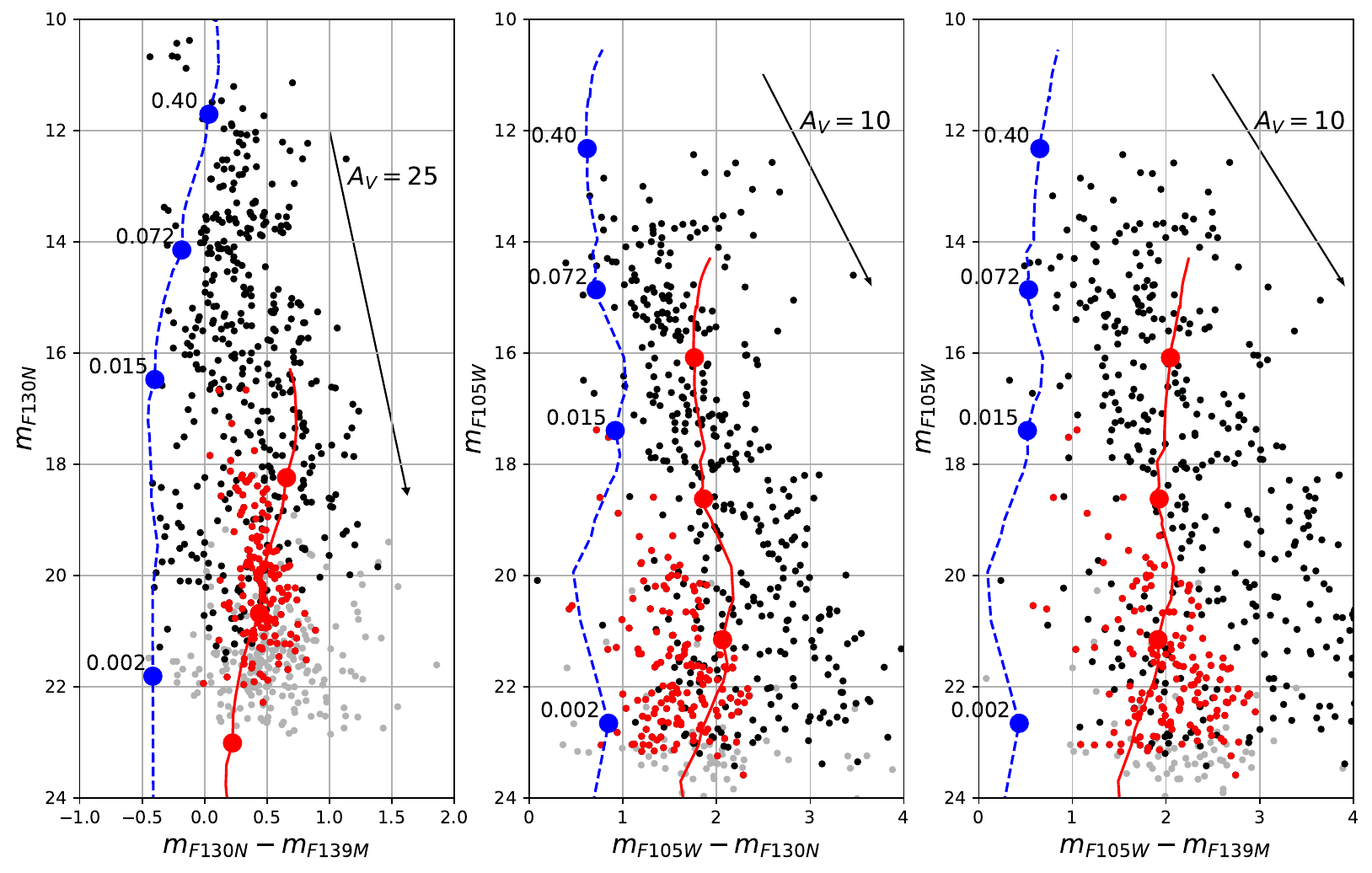}
\caption{CMDs for all sources in our catalog with uncertainty $dm \leq 0.2$~mag in all filters (black dots). Sources with larger uncertainty up to $dm=0.3$~mag, { generally at the bottom of each diagram,} are shown as gray dots. { The red dots indicate candidate background sources selected according to the diagram shown in Figure~\ref{fig:Av_Analysis}.} The blue dashed line is the 1 Myr isochrone according to the BT-Settl models, { modified in the $F139M$ passband as in \cite{Robberto-ONC-2020ApJ...896...79R}.} The red solid line shows the same isochrone with $A_{V}$=10 magnitudes of reddening according to the standard \cite{Cardelli_1989ApJ...345..245C} law with $R_V=3.1$, the corresponding vector being represented by an arrow. The filled circles on the isochrones, top-to-bottom, indicate the locus of a 0.4, 0.72, 0.015 and 0.002~\Msun~ object.}
\label{fig:3CMDs}
\end{figure*}

{ Guided by the reddening vectors, one can discern that the main  distribution of sources appears compatible with a heavily reddened IMF, peaking in the mass bin 0.075-0.40~\Msun\, and spread to higher magnitudes and redder colors, i.e. bottom-right. If one compares the three diagrams,  however, an anomaly becomes apparent: the clumps of points in the lower half of the diagrams cannot be reconciled with the isochrone assuming a similar amount of extinction. In fact, in order to match the highest density of sources, the three reddened isochrones, represented in the plots as red solid lines, had to be traced assuming $A_V=25$ in the first plot and $A_V=10$ in both the second and third plot . The strong discrepancy in $A_V$ values means those sources cannot be simply interpreted as reddened young stars, but represent instead the population of background galactic sources ``leaking'' through the nebular background. Comparing the first CMD with the similar one presented by  \cite{Robberto-ONC-2020ApJ...896...79R}, one can see that the locus of this population is consistent with what already found for the ONC.

To discriminate more rigorously the cluster sources from the contaminants, we have looked at the discrepancy between the extinction values. Dereddening each source to the 1~Myr isochrone one determines three different $A_V$ values, one for each plot. Plotting the largest (max-min) vs  minimum difference (central-min) of each $A_V$ trio, one can draw a region encompassing sources that present a large outlier, i.e with the first difference much larger than the second one, see Figure~\ref{fig:Av_Analysis}. This results in 179 sources concentrated in the lower parts of the CMDs (red dots in Figure~\ref{fig:3CMDs}, that we shall regard as candidates galactic contaminants. Consequently, we are left with with 371 candidates cluster sources. }

\begin{figure*}
\centering
\includegraphics[width=0.45\textwidth]{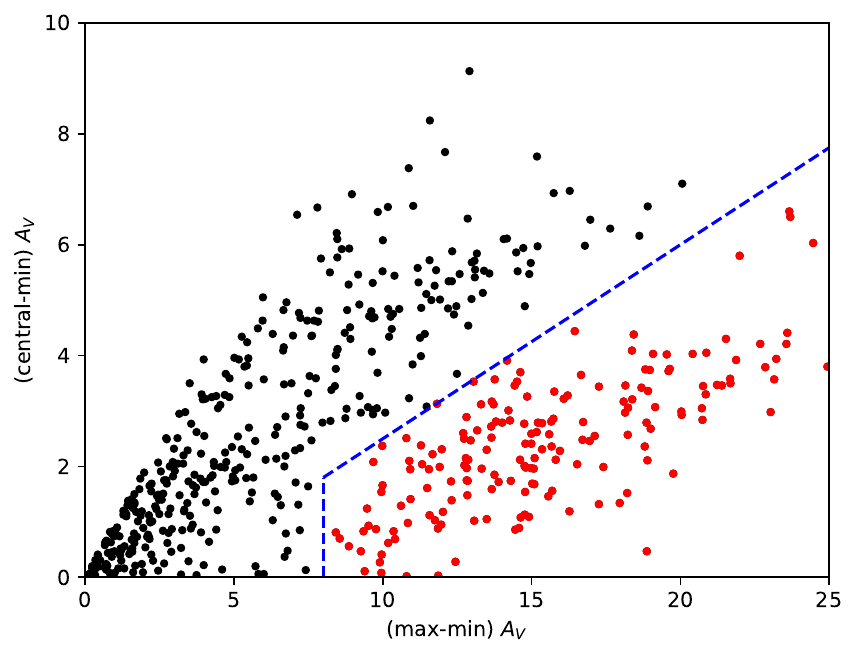}
\caption{ Largest vs. smallest difference between the three $A_V$ values determined by dereddening each source to the 1~Myr isochrones in Figure~\ref{fig:3CMDs}. Red dots indicate the sourses showing the largest discrepancy (i.e. a large outlier), that can be considered as candidate bagkground sources }
\label{fig:Av_Analysis}
\end{figure*}

{ }
Returning our attention the first CMD of Figure~\ref{fig:3CMDs}, one can see that none of these candidate background sources has negative H$_2$O index. All the faint objects with negative H$_2$O color index lying parallel to the blue isochrones remain identified as candidate cluster stars. They most probably represent the fraction of very low mass sources, down to planetary masses, detected by our survey. Their location in the diagram is consistent with that of the objects detected in the ONC, albeit in lower number due to the large and non-uniform extinction across the field and the more modest statistics for this less rich cluster, considering also our smaller survey area. 
}

\par\vfill\eject

\subsection{Color-Color Diagram and Extinction}\label{sec:2CD}
\begin{figure*}
\centering
\includegraphics[width=0.55\textwidth]{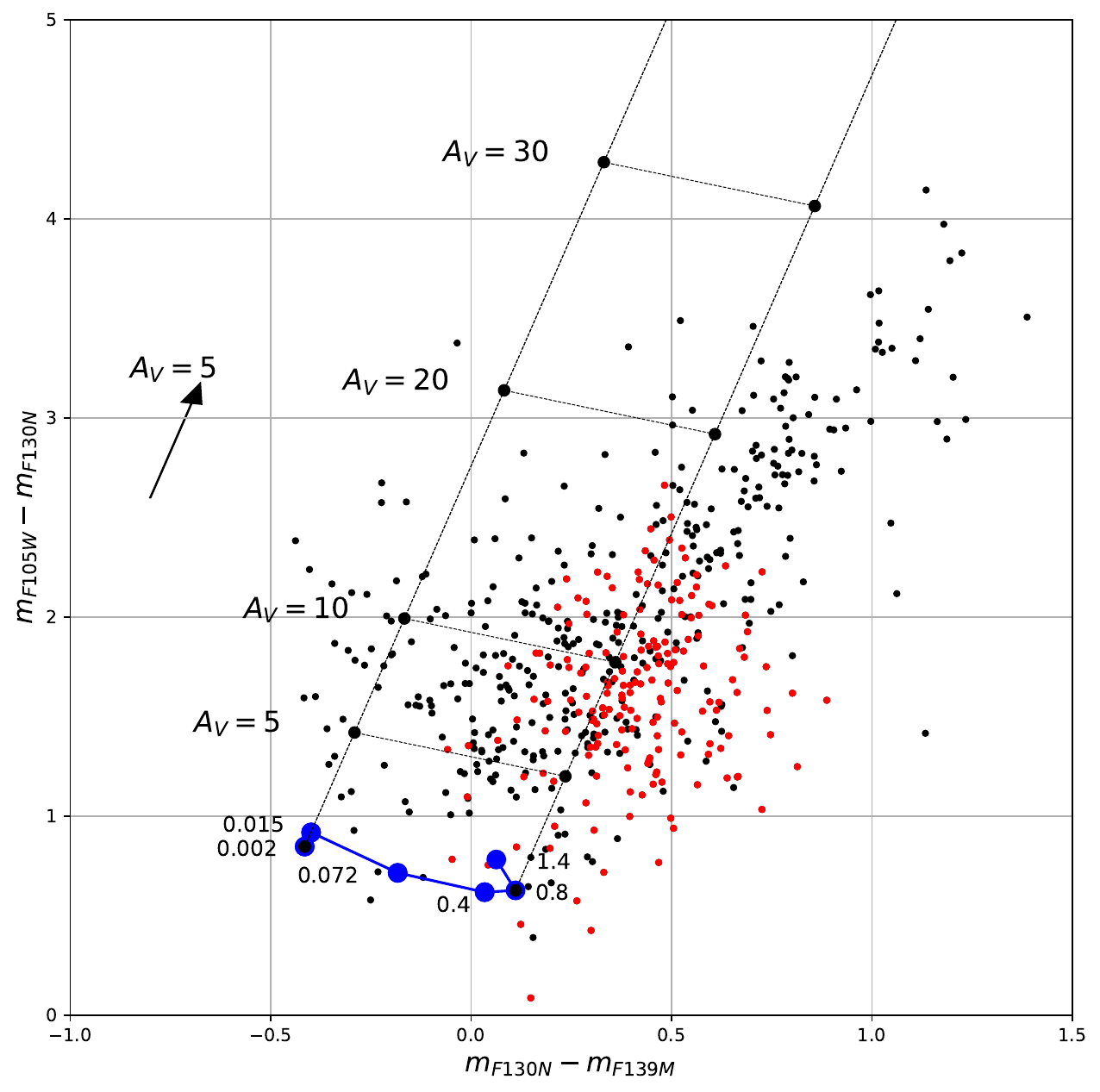}
\caption{ Color-Color diagram for the sample of sources with magnitude error smaller than 0.2 in all three filters. The blue solid line represents the locus of the 1 Myr BT-Settl isochrone, with large blue dots drawn in correspondence of the masses values reported in units of solar mass. The black dotted straight lines are oriented along a standard $R_V=3.1$ reddening vector \citep{Cardelli_1989ApJ...345..245C}, with reddening corresponding to $A_V =5, 10, 20, 30$ magnitudes. The red dots indicate our candidate background sources. }
\label{fig:CCD_Linear}
\end{figure*}

The availability of a third photometric band allows creating a distance independent (F130N-F139M) vs. (F105W-F130N) color-color diagram, shown in Figure \ref{fig:CCD_Linear}, for the subset of 550 targets having uncertainties smaller than 0.2~mag in all three filters. The 1~Myr isochrone, drawn as a blue solid line for masses ranging from 1.4~\Msun\ to 0.002~\Msun, spans a rather narrow range of colors, as expected due to the nearly adjacent wavelengths; { note the negative values of the (F130N-F139M) index for $M<0.4~\Msun$\, due to the H$_2$O absorption feature.} 

Overall, the source distribution is dominated by extinction, spreading the points up and to the right toward reddened colors.  The two straight  lines trace the reddening corresponding to
the \cite{Cardelli_1989ApJ...345..245C} $R=3.1$ extinction law, estimated at the nominal filter wavelengths of 1.05\um, 1.30\um, { and 1.39\um, } as explained in the previous section. Assuming our passbands are not significantly contaminated by non-photospheric emission, these lines { should} enclose the area occupied by reddened low-mass cluster members.  { Note, however, that our candidate cluster (black dots) and background sources (red dots) largely overlap in the diagram, which therefore cannot be used to further discriminate the two populations. }

It is also clear that the straight lines fail to explain the main distribution of sources, { as they are spread along a direction less steep than that predicted by our reddening vector.}
We have verified that other reddening laws { trace straight lines that are not significantly different} from the  \cite{Cardelli_1989ApJ...345..245C} vector, as expected due to the limited spectral coverage of our passbands { and the relative uniformity of reddening laws in the near-IR.}

\begin{figure*}
\centering
\includegraphics[width=0.95\textwidth]{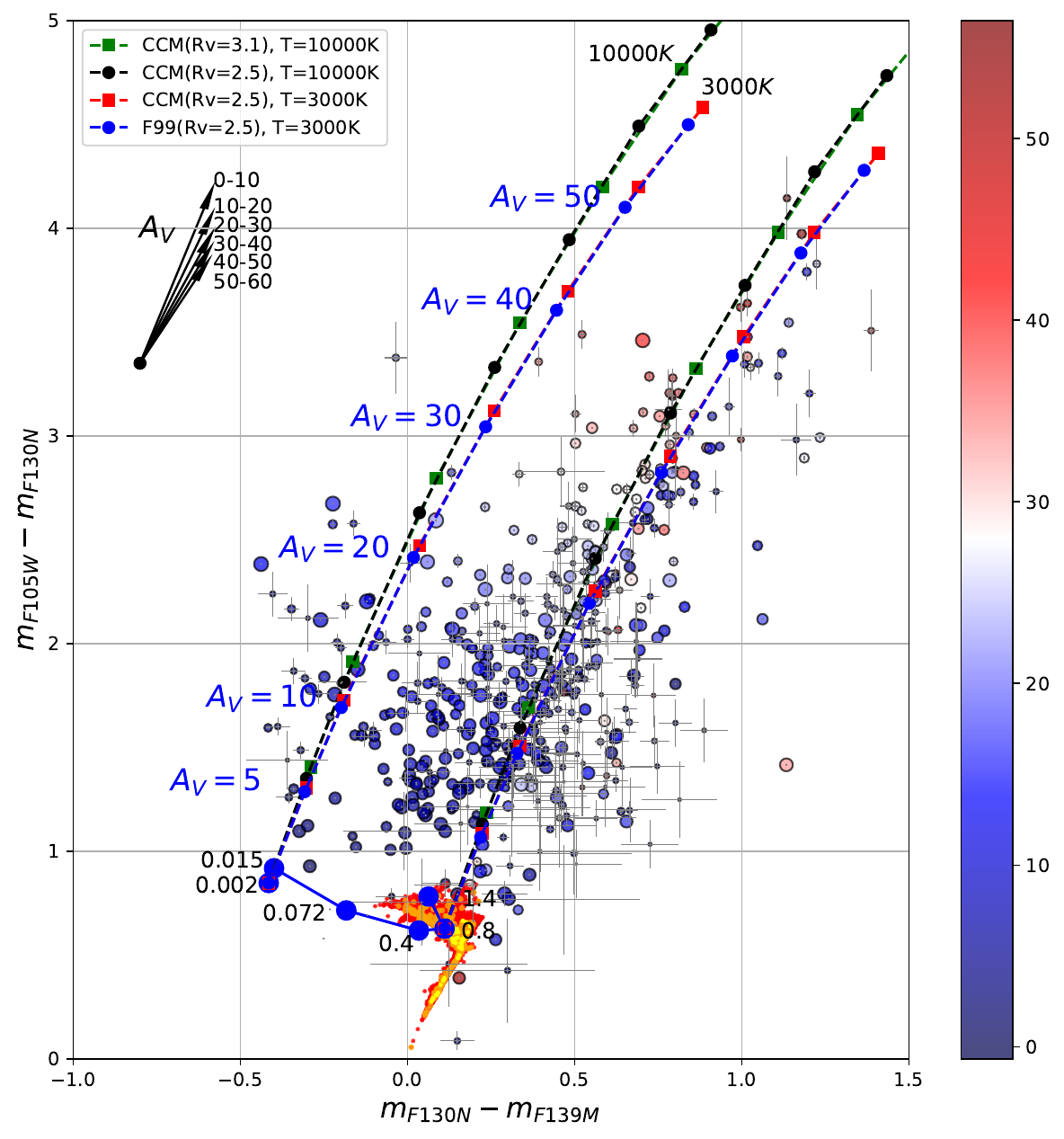}
\caption{Color-Color diagram for the sample of sources with magnitude error smaller than 0.2 in all three filters. The blue solid line represents the locus of the 1 Myr BT-Settl isochrone, with the correction for the F139M filter adopted by \cite{Robberto-ONC-2020ApJ...896...79R} and filled circles drawn in correspondence of the masses reported in units of solar mass. The  dashed lines represent the reddening curves determined through synthetic photometry, according to the parameters reported in the labels (see also text). On each curve the dots represent, from the isochrone up, the reddening corresponding to $A_V =5, 10, 20, 30, 40, 50$ and $60$ magnitudes. The arrows represent the change of reddening vector for increasing ranges of extinction.  The  size of the circles is proportional to the magnitude in the F130N filter, whereas the colors are representative of the extinction according to the \cite{Fitzpatrick_1999PASP..111...63F} $R_v=2.5$ reddening law, with color scheme shown in the scalebar to the right. 
{ All error bars are plotted, derived from the uncertainties presented in Figure~\ref{fig:dmag_vs_mag}, but only those larger than the circle size are visible. 
{ The clump of colored dots (red, orange, and yellow colors tracing increasing  source density)} at the bottom of the figure represents the tip of the galactic population derived from the Besan\c{c}on models (Robin et al. 2003) for the celestial coordinates and extent of our NGC2-2024 field. See Section~\ref{sec:2CD} for details.}
}
\label{fig:CCD_Avcoded}
\end{figure*}

The mismatch between { the standard reddening vectors} and the source distribution { in our diagram} can be reconciled if one observes that { in the case of NGC~2024} we are dealing with extremely high values of extinction. For any given reddening law, as extinction increases { and the the spectral energy distribution becomes redder and redder}, one has to account for the change of effective wavelength within the passbands. { The effect is more pronounced in broad-band filters and causes significant non-linearity if the color is obtained combining narrow and broad passbands, as the effective wavelengths remain nearly constant in the first one and moves significantly in the second one. Moreover, there is a tendency to reach saturation, i.e. as the extinction increases the incremental variations of effective wavelengths become smaller and smaller. Also, the estimated change of source magnitude in each filter depends on the original spectral energy distribution of the source itself. }

To assess the relevance of these factors, we have performed synthetic photometry in our photometric passbands probing different assumptions. Specifically, we have determined the change of our colors at   $A_V=5,10,20,30,40,50,60,$ for effective temperatures $T_{eff}=10,000$~K (a reference { value} for the Vegamag scale) and $T_{eff}=3,000K$ (more representative { of low mass young stars}) and for different reddening laws. The results are always  a set of nonlinear relations { like those} presented in Figure~\ref{fig:CCD_Avcoded}, that generally better match the locus of the sources. The curves shown in Figure~\ref{fig:CCD_Avcoded}, plotted twice for the low-mass and high-mass end of our 1~Myr isochrone, are relative to four cases. There is a degeneracy, as one can readily discern only two curves { departing from the same points}: the leftmost is relative to two cases with $T_{eff}=10000$~K while the rightmost is for two cases at $T_{eff}=3000$~K. { Their difference  illustrates how the change in colors depends on the effective temperature of the star, colder stars producing stronger deviations from the original slope, i.e. the tangent at  $A_V$=0}. 

The two cases calculated for $T_{eff}=10000$~K probe two different $R_V$ values for the same \cite{Cardelli_1989ApJ...345..245C} law: R$_V$=3.1 
(green squares) and $R_v=2.5$ (black circles). { We consider this second value following \cite{Damineli_2016MNRAS.463.2653D}, who found that intracluster dust grains have properties different from the lower density ISM, resulting in a reddening law steeper than the canonical one. They report a value  $R_v=2.5$ for the young stellar cluster Westerlund 1, one of the most massive young clusters in the Milky Way. While the two curves overlap, the lower $R_V$ value 
produces higher $A_V$ for the same color excess: black circles are always below the green squares. The larger estimated extinction resulting from the adoption of a lower $R_V$ will lead to an increase} of the estimated source luminosity, photospheric radius, and a younger isochronal age.

The other curve, relative to a 3000~K photosphere, is also the overlap of two different cases, two different families of reddening laws for the same $R_V=2.5$ value: the \cite{Cardelli_1989ApJ...345..245C} (red squares) and the \cite{Fitzpatrick_1999PASP..111...63F} (black circles). The differences in this case are smaller, with the 
\cite{Fitzpatrick_1999PASP..111...63F} law predicting slightly higher $A_v$ values for the same amount of reddening.



The extinction at  near-IR wavelengths, $0.8<\lambda<2.4~\um$ is generally parameterized by a power law $A_\lambda\propto \lambda^{-\alpha}$ \citep{Fitzpatrick_1999PASP..111...63F}. As reported by \cite{Damineli_2016MNRAS.463.2653D}, the power-law exponent $\alpha$ can take a broad range of values,  from 1.66 for the \cite{Cardelli_1989ApJ...345..245C} $R=3.1$ reddening curve to $\alpha=2.13$  for Westerlund 1, up to $\alpha=2.64$ \citep{Gonzalez-Fernandez_2014ApJ...782...86G}. 
In our case the spectral coverage is admittedly narrow, but it is still interesting to check how $\alpha$ changes in our four reference cases performing a best fit each tern of extinction values. In order to maintain the same parameterization adopted by \cite{Damineli_2016MNRAS.463.2653D}, we add to our synthetic estimates the $K_s$ filter adopting the 2MASS passband. It results that, for extinction values ranging from $A_V=1$ to $A_V=60$, $\alpha$ takes the following values:
\begin{itemize}
\item
$\alpha = 1.62 - 1.78$ for \cite{Cardelli_1989ApJ...345..245C} with $R_V=3.1$ and $T_{eff}$=10000~K
\item
$\alpha = 1.62 - 1.73$ for \cite{Cardelli_1989ApJ...345..245C} with $R_V=2.5$ and $T_{eff}$=10000~K
\item
$\alpha = 1.62 - 1.77$ for \cite{Cardelli_1989ApJ...345..245C} with $R_V=2.5$ and $T_{eff}$=3000~K
\item
$\alpha = 1.50 - 1.61$ for \cite{Fitzpatrick_1999PASP..111...63F} with $R_V=2.5$ and $T_{eff}$=3000~K
\end{itemize}
Overall the change is of the order of 10\%. While the $\alpha$ values always increases with the extinction, the change is not enough, for our passbands, to move it to values as high as e.g.  $\alpha$=2.13 with $A_V\sim10$ estimated by  \cite{Damineli_2016MNRAS.463.2653D} for Westerlund 1 on the basis of broad-band multi-color photometry. 

{ In summary, as reddening increases extinction increases non-linearly. It takes a larger amount of extinction to produce the same increment of reddening or, in other terms,  the $R_\lambda = A_\lambda / (A_{\lambda_{x}} - A_\lambda)$ ratio decreases with $A_\lambda$. The correlation between $R_\lambda$ and A$_V$, is illustrated by the set of arrows in Figure~\ref{fig:CCD_Avcoded}, representing how the reddening vector changes in our color-color diagram, both direction and module, when one increases $A_V$ in steps of 10 magnitudes. 
We shall consider hereafter two reference case: the standard \cite{Cardelli_1989ApJ...345..245C} $R_V=3.1$ law for $T_{eff}=10,000K$ and the   \cite{Fitzpatrick_1999PASP..111...63F} $R_V=2.5$ law for $T_{eff}=3,000K$, anticipating that the two assumptions will result in different parameters for the cluster population. 

Adopting these laws, we obtain two estimates of the extinction toward each source using the following strategy. Tracing back the reddening curve, we deredden to the 1~Myr isochrone all cluster candidates lying in the central region between the two reddening curves drawn in Figure~\ref{fig:CCD_Avcoded}.
Cluster candidates to the left and right of the central band require an approximate treatment, since there is no reddening curve, among those that we have considered and that correspond to a rather broad range of models, that can consistently explain their location on the diagram.
We thus consider only their F105-F130 color, as it combines two bandpasses centered on the photospheric continuum and, by producing the largest differences, is the one less affected by photometric uncertainties. To account for the intrinsic stellar color, we subtract from the  measured F105-F130 indexes the color of a 0.8~\Msun\, star on our isochrone, almost coincident with the color of a 0.4~\Msun star.  We verified that for the sources within the central region this approach produces A$_V$ values close to those obtained with the full projection back to the isochrone. 
The same strategy is adopted for sources to the left of the central band, subtracting this time the color of a 0.015\Msun\, dwarf.

{ The clump of colored dots} at the bottom of Figure~\ref{fig:CCD_Avcoded} represents the tip of the galactic population derived from the Besan\c{c}on models \citep{Robin+2003A&A...409..523R} for the celestial coordinates and extent of our NGC2~2024 field. Running the Besan\c{c}on simulator, we have maintained the default parameters removing cuts on distance and apparent magnitudes that may bias the results, resulting in full sample of about 22,500 stars. The photometry in our bandpasses was derived assuming for each model star the Phoenix spectrum \citep{2013A&A...553A...6H} appropriate for the given effective temperature and surface gravity and accounting for the stellar luminosity and distance from the Sun. 
Considering that the locus of a 0.8~\Msun\, star is nearly coincident with the peak of the Besan\c{c}on distribution (the tip of the yellow area in Figure~\ref{fig:CCD_Avcoded}) we de-project to this point all background candidate sources, regardles on their position in the diagram. For those lying withing the central band, the difference between the extinction values determined using this treatment or a full deprojection to the isochrone is generally small.

Producing  Figure~\ref{fig:CCD_Avcoded},} we have color coded each source according to the extinction values derived as we just described using the \cite{Fitzpatrick_1999PASP..111...63F} law. A color code based on the \cite{Cardelli_1989ApJ...345..245C} law would produce almost exactly the same diagram, with the exception of the scale bar with maximum at A$_V$ = 42.  

\begin{figure}
\centering
\plottwo{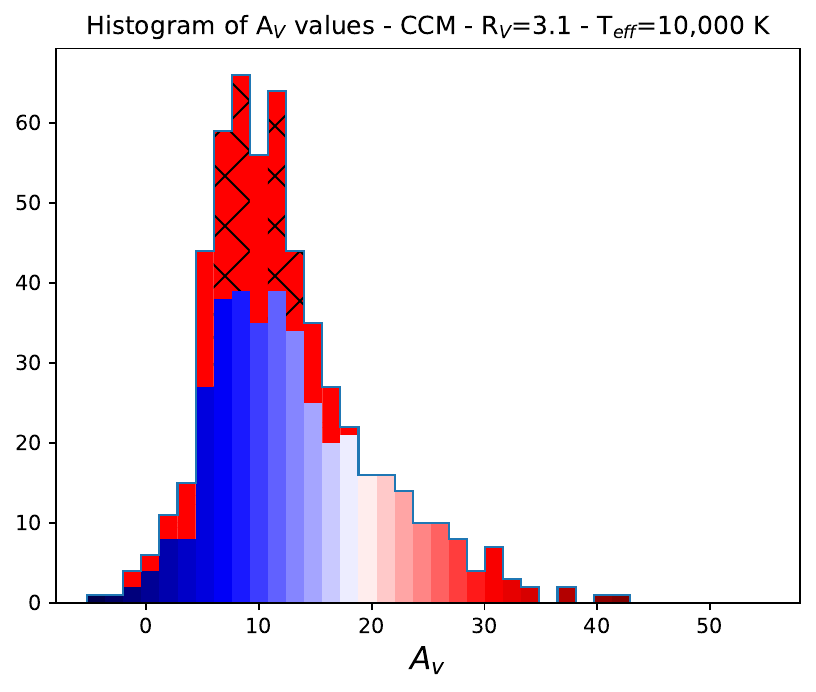}{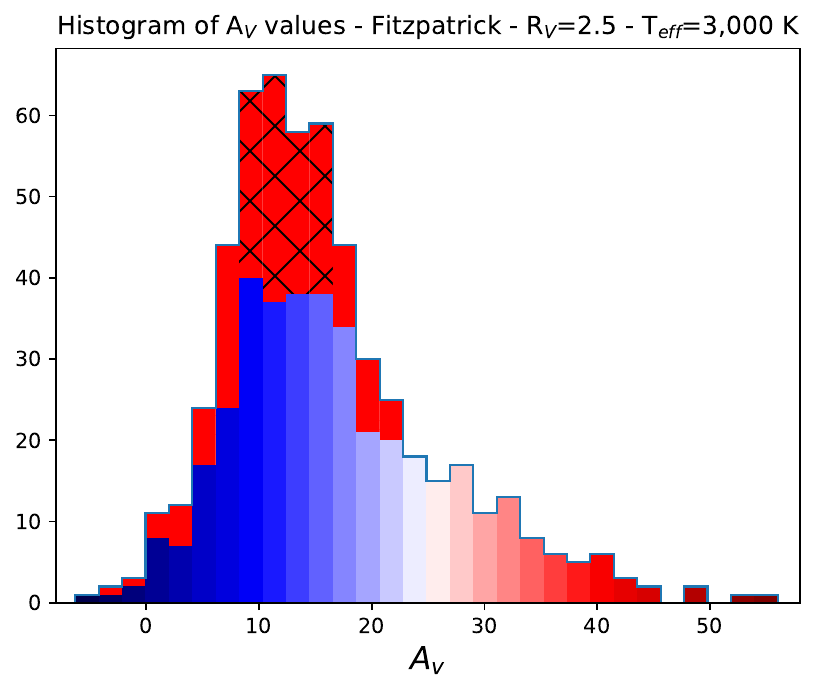}
\caption{Histogram of the distribution of estimated $A_V$ values according to { the \cite{Cardelli_1989ApJ...345..245C}   (left) and \cite{Fitzpatrick_1999PASP..111...63F}  (right) reddening curves.  For the cluster candidates the  $A_V$ values are color-coded according to the scalebars to the right. The cross-hatched area shows the background candidate sources, plotted on-top of the cluster candidates.}}
\label{fig:Av_2_histograms}
\end{figure}

\begin{figure}
\centering
\includegraphics[width=0.45\textwidth]{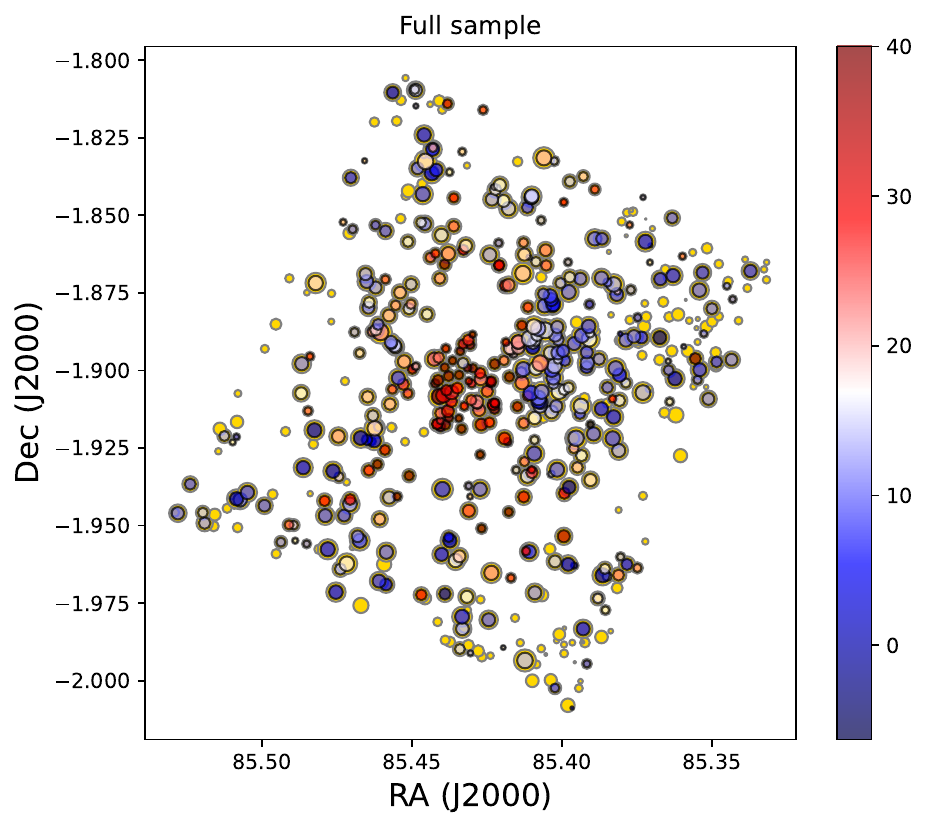}
\caption{ Map of the detected sources color coded according to $A_V$. The scale bar is relative to the \cite{Fitzpatrick_1999PASP..111...63F} law.}
\label{fig:Av_map_REV2}
\end{figure}

In Figure~\ref{fig:Av_2_histograms} we present the histograms of the $A_V$ distribution. { The candidate cluster members have been color coded using the same scale adopted for Figure~\ref{fig:CCD_Avcoded}, whereas the candidate background sources  have been plotted in yellow {\sl on top} of the cluster distribution. The two plots are similar, with the \cite{Fitzpatrick_1999PASP..111...63F} law producing higher A$_V$ values. In each plot the maximum is about the same for both the cluster and background populations. There are no highly reddened background sources, arguably a selection effect as we are not considering the faintest sources with large photometric error (i.e. they gray dots at the bottom of the diagrams in Figure~\ref{fig:3CMDs}), that in the CMDs generally fall in the clump of contaminants.

Figure~\ref{fig:Av_map_REV2} shows the spatial distribution of the sources, color coded according to the \cite{Fitzpatrick_1999PASP..111...63F} law. Again, a nearly identical plot can be obtained using \cite{Cardelli_1989ApJ...345..245C} law, with the exception of the compressed scale bar. 
The map} confirms that the most highly reddened sources are concentrated at the center, { while} the sources with modest reddening are concentrated to the West, consistent with previous findings that the region to the West of the dark filament is the one less obscured and possibly more evolved. { The candidate background sources are scattered around, with  some clustering at the southern and western corners of the field, suggesting that these are regions where the underlying molecular clouds has lower optical depth.}

\subsection{Luminosity Function}\label{sec:LuminosityFunction}
\begin{figure*}
\centering
\plottwo{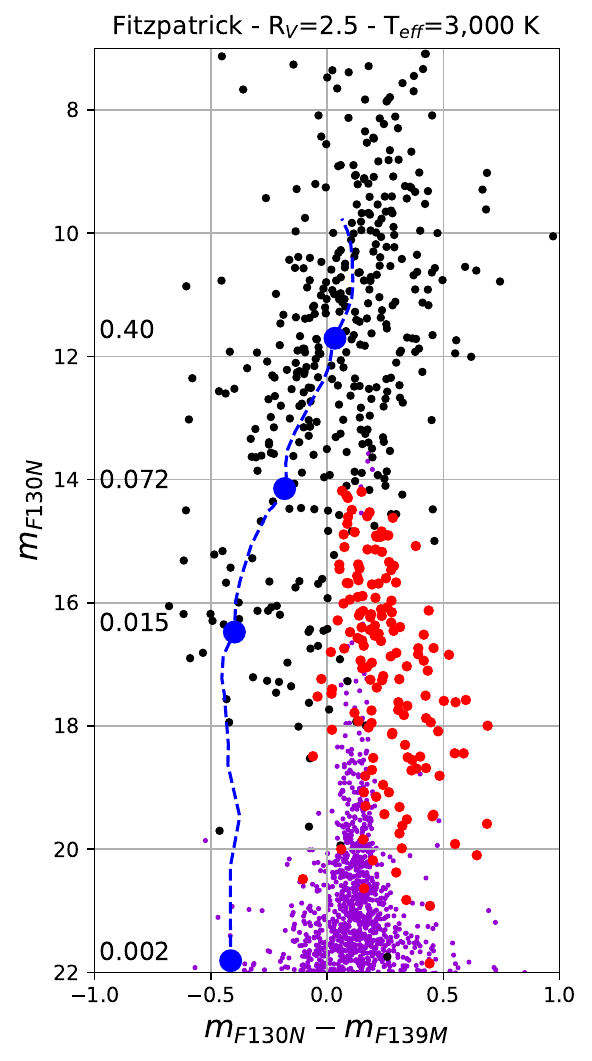}{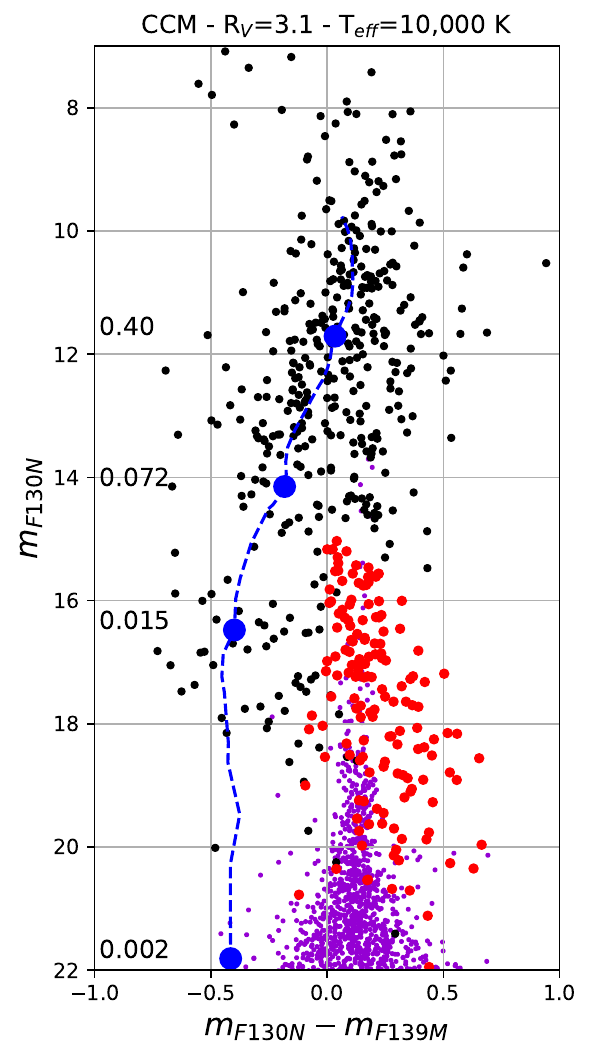}
\caption{CMDs of the dereddened sources with magnitude error smaller than 0.2 in all three filters, after applying the \cite{Fitzpatrick_1999PASP..111...63F}-$R_V=2.5$ (left) and \cite{Cardelli_1989ApJ...345..245C}-$R_V=3.5$ (right) reddening laws. The blue dashed line represents the locus of the 1 Myr BT-Settl isochrone, with the correction for the F139M filter adopted by \cite{Robberto-ONC-2020ApJ...896...79R}. Filled circles mark the masses reported on the left axis, in units of solar mass. 
The red circles indicate the background candidate sources, while the purple dots
mark the full sample of sources simulated by the Besan\c{c}on model converted to our filters using the Phonix models. 
}. 
\label{fig:CMD_parabola_Levine}
\end{figure*}

Using the estimated extinction values, one can produce dereddened versions of the original CMDs presented in Figure~\ref{fig:3CMDs}. { In particular}, in Figure~\ref{fig:CMD_parabola_Levine} we show the {dereddened} F130N vs. F130N-F139M CMDs, { for our two reference extinction laws.  Each diagram shows again the BT-Settl 1~Myr isochrone (blue dashed line) with the four large dots indicating the locus of a 0.4, 0.072, 0.015 and 0.002 \Msun\, object. We distinguish between the NGC~2024 cluster candidate population (black dots), dominating the upper part of the diagram, and the sources classified as candidates background stars (blue dots), concentrated in the lower part of the diagram. In both diagrams the two groups appear well separated, with few interlopers at the fringes of their distributions. 
Using the \cite{Fitzpatrick_1999PASP..111...63F} law, the dereddened magnitudes of all sources result brighter by a couple of magnitudes than those derived using the \cite{Fitzpatrick_1999PASP..111...63F} law. The F130N-F139M colors instead are similarly distributed, and for the cluster candidates they become increasingly negative going to fainter magnitudes. Also this diagram thus confirms the presence of H$_2$O in absorption in low mass objects. }

\begin{figure}
\centering
\plottwo{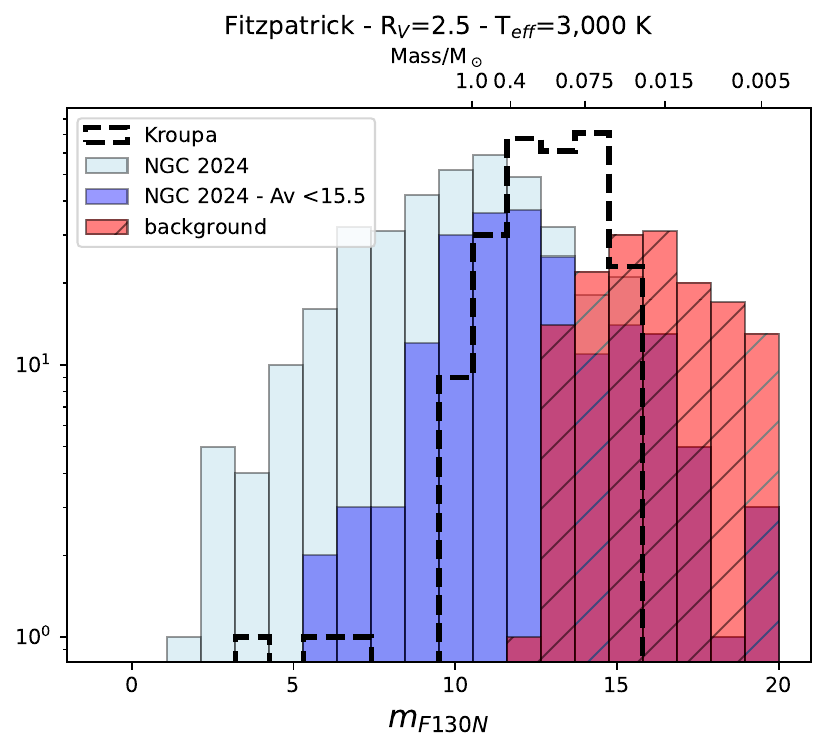}{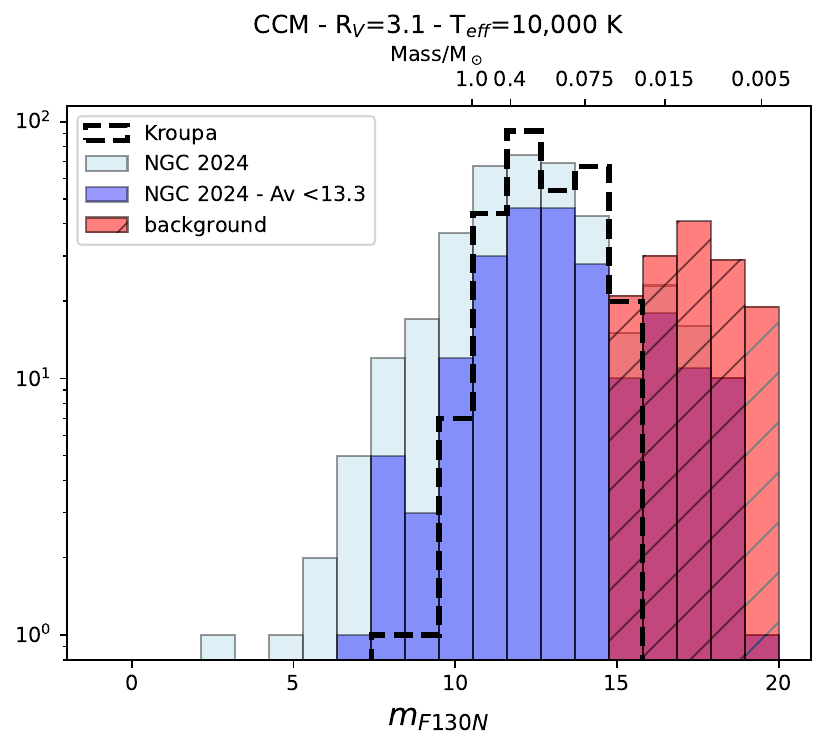}
\caption{F130N luminosity functions for the { candidate} cluster members, both full sample (light blue) and extinction limited sample (dark blue), together with the background sources (red dashed area). The top axis shows the masses corresponding to the F130N magnitudes according to the 1~Myr BT-Settle model, for an assumed distance of 400~pc. A random realization of the Kroupa \citep{Kroupa_2001MNRAS.322..231K} IMF for the number of stars of the extinction-limited sample is also plotted as a dotted line. The left plot is relative to the masses estimated using our baseline reddening law \citep[][with $R_V=2.5$ and $T_{eff}=3000$~K,]{Fitzpatrick_1999PASP..111...63F}  whereas the plot to the right is relative to the reddening curve derived from \cite{Cardelli_1989ApJ...345..245C} with $R_V=2.5$ and $T_{eff}=10000$~K.}
\label{fig:F105W_LuminosityFunction}
\end{figure}

{
As in Figure~\ref{fig:CCD_Avcoded}, we use colored dots to represent the Galactic population estimated from the Besan\c{c}on model of the Milky Way. Plotting these points, we added to the synthetic photometry of each star an error randomly extracted from the histograms presented in Figure~\ref{fig:3magnitudeplots}, considering only sources in a $\pm0.1$ wide bin centered around the source's magnitude. 
The model predicts for the few, brightest Besan\c{c}on sources $m_{F130N}\simeq14$, whereas our candidate sources are about 1 magnitude brighter and fainter for the  \cite{Fitzpatrick_1999PASP..111...63F} and \cite{Cardelli_1989ApJ...345..245C} laws, respectively. Colors are more spread than those of the Besan\c{c}on population. One has to consider that photometric uncertainties directly affect not only the measured colors, an effect we have accounted for by spreading the Besan\c{c}on data, but also the estimated extinction derived from Figure~\ref{fig:CCD_Avcoded}, sensitive to colors. This is a source of uncertainty we did not account for. Note also that the blue dots appear more scattered on the right side of the Besan\c{c}on distribution, for both extinction laws. Having a  fraction of sources with some residual redder color may suggest that when we assumed for all of them the intrinsic colors of a star at the tip, rather than the tail, of the yellow area in Figure~\ref{fig:CCD_Avcoded}, we may have occasionally underestimated extinction.

The histograms of the dereddened F130N magnitudes are shown in Figure~\ref{fig:F105W_LuminosityFunction}. 
Plotting the values for the cluster candidates, we distinguish between the full set of 397 sources  (light blue) and an extinction-limited sample (dark blue) containing the 50\% of sources with lower extinction. We also plot the background contaminants (red dashed). The bins are equally spaced in magnitude by an amount corresponding to a change of brightness by a factor of 2.   
The scale at the top of the plot also shows the mass values according to the 1~Myr BT-Settl isochrone at 400~pc distance, obviously applicable only to the NGC~2024 sample. 

As explained in Section~\ref{sec:observations}, we cannot perform a reliable completeness correction at the faint end. We can  focus instead on the brighter stars, i.e. the left side of the histograms free from background contamination. 
Referring for convenience to the mass scale, the IMF resulting from the \cite{Fitzpatrick_1999PASP..111...63F} law appears heavily loaded toward high masses compared with a conventional distribution like e.g. the realization of the Kroupa IMF \citep{Kroupa_2001MNRAS.322..231K} also shown in both figures as a dashed line. The extinction limited sample also presents the same peak shifted to masses higher than the Kroupa IMF, but with a steeper slope than the full sample. The  \cite{Cardelli_1989ApJ...345..245C} law (right plot) produces instead an IMF in better agreement with the Kroupa IMF, both the full sample and especially the extinction limited sample. Comparing the two diagrams, we can conclude that that the \cite{Fitzpatrick_1999PASP..111...63F} law with $R_V$=2.5 for $T_{eff} = 3000~K$ does not seem applicable to the case of NGC~2024.

Still, also with \cite{Cardelli_1989ApJ...345..245C} law the there is an apparent excess of high luminosity sources. A Wilcoxon rank-sum test shows a $\simeq 10$\% probability that the full sample can be randomly extracted from a Kroupa IMF. The fact that the excess is more evident for the full sample { points to a correlation} between extinction and luminosity. The presence of such a correlation can be expected, since any overestimate of the extinction naturally leads to an overestimate of the intrinsic stellar luminosity, and viceversa. In this case the effect must be systematic, and looking at the two CMDs in Figure~\ref{fig:3CMDs} one may discern a higher density of dots along a line parallel and above to the 1~Myr isochrone. This clumping is more clear in the \cite{Fitzpatrick_1999PASP..111...63F} diagram between $m_{130} = 10$ and 13, about two magnitudes above the isochrone, but it may still be visible on the right diagram, with larger scatter. This could indicate that the 1~Myr isochrone, the youngest available for the BT-Settl class of models, may not be ideal for a very young,  deeply embedded cluster. The older \cite{D'Antona&Mazzitelli-1994ApJS...90..467D} models pushed the calculations to earlier times, showing that the luminosity of a 0.4~\Msun\, star decreases by about 50\% as it evolves from 0.5 to 1~Myr. 

Of course, positing younger ages for the stars in NGC~2024, one should take into account that mass accretion rates should also be higher, contributing to both the stellar luminosity and  the actual pre-main-sequence evolution \cite[see e.g.][for a discussion]{ToutLivioBonnell_1999MNRAS.310..360T}. 
Future JWST observations aiming at the determining the spectral type, surface gravity and reddening of each source will shed light on the cause of the apparent excess of high luminosity stars in NGC~2024 and in general on the IMF of this extremely young and deeply embedded cluster.

}

\begin{deluxetable}{ccc}
\tablewidth{6cm}
\tablecaption{Extinction and Mass of { candidate} cluster members\label{tab:AV_M}}
\tablehead{
\colhead{Entry Nr.} & \colhead{$A_V$}& \colhead{$M/\Msun$}}
\startdata
1  & 4.75 & 0.04 \\
3  & 6.70 & 0.014\\
8  & 6.89 & 0.008\\
\ldots & \ldots & \ldots\\
\enddata
\tablecomments{ Values estimated adopting the \cite{Cardelli_1989ApJ...345..245C} $R_V=3.1$ reddening law. The full table is available for download in the electronic version of the paper.}
\end{deluxetable}

\section{Comparison with IR spectroscopy from Levine et al. (2006)}\label{sec:Levine}
In their near-infrared photometric and spectroscopic study of NGC~2024 \cite{Levine+2006} classified about 70 sources using the $J$-and $H$-band water absorption features, deriving spectral types in the range $\sim$ M1 to later than M8. { Performing ground-based spectroscopy,} the classification of infrared spectra { based} on the actual depth and width of the 1.4~$\mu$m H$_2$O feature { is possible only for the brightest sources, as telluric absorption allows measuring only }the slope and depth of the short-wavelength falloff at 1.35~$\mu$m. Still, combining this information with that coming from the shape of the 1.68~$\mu$m feature in the $H$-band, \cite{Levine+2006} managed to classify their sample with typical errors of 0.5 to 1 sub-types. This provides us with an independent test of our photometric classification method. 

\begin{figure*}
\centering
\plottwo{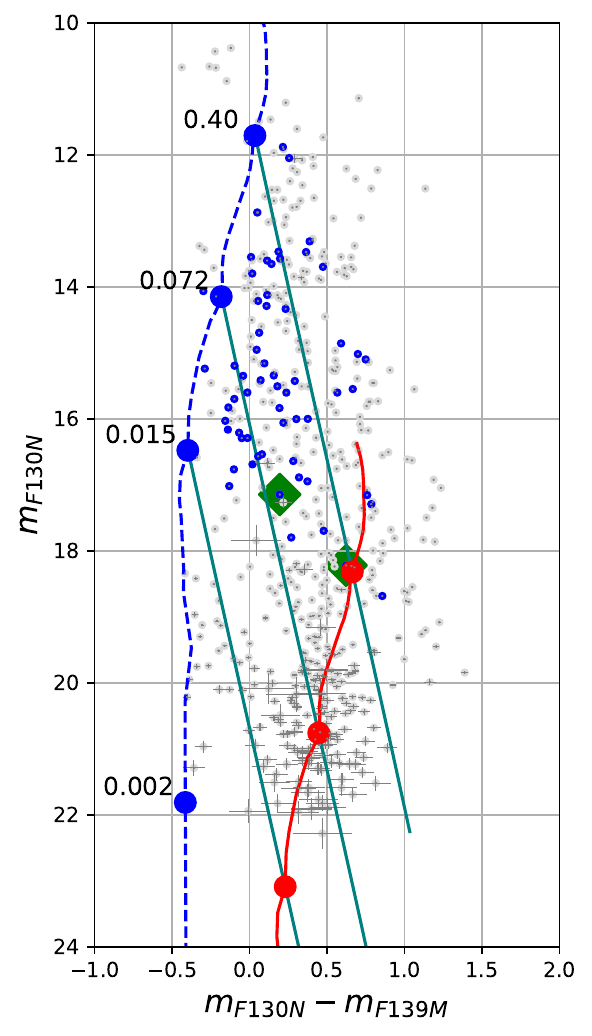}{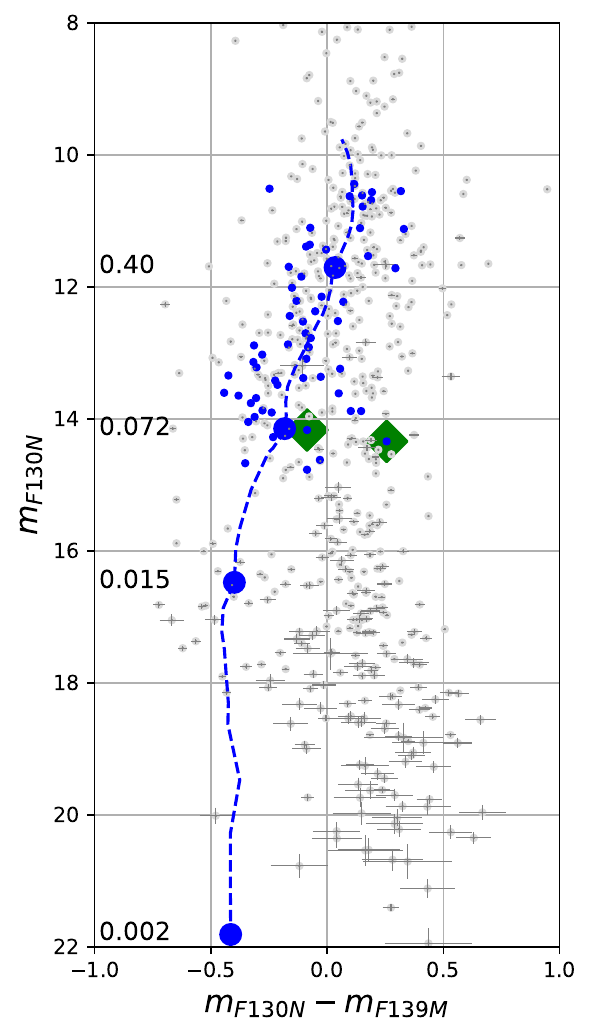}
\caption{ Left: The F130N vs. F130N-F139M CMD diagram presented in Figure~\ref{fig:3CMDs}-left, relative to the sources before reddening correction, is shown here for the full sample (gray dots). Blue dots indicate the sources targeted by the spectroscopic survey of \cite{Levine+2006}. Green diamonds represent the two sources classified by \cite{Levine+2006} as background M-giants.The green solid lines trace the extinction curves according to the \cite{Cardelli_1989ApJ...345..245C} reddening law, drawn from the positions of a 0.4, 0.072 and 0.015 \Msun object on the 1~Myr isochrone without extinction (blue dashed line) to the same isochrone with $A_v$=15 . Right: The same diagram after reddening correction (see also the right diagram in Figure~\ref{fig:CMD_parabola_Levine}), with blue dots indicating the sources targeted by the spectroscopic survey of \cite{Levine+2006}. The green diamonds represent again the two sources classified by \cite{Levine+2006} as background M-giants. }
\label{fig:CMD_Levine}
\end{figure*}

Figure~\ref{fig:CMD_Levine}-left shows the same data { presented} in Figure~\ref{fig:3CMDs}-left, i.e. the {\sl observed} magnitude { and color}, with the sources classified by \cite{Levine+2006} indicated as red dots. They occupy the upper part of the diagram, as expected due to the { significant} difference in sensitivity between ground-based spectroscopy and HST imaging. Analyzing their sample, \cite{Levine+2006} identified two sources (\#60 and \#64) that may be background M giants, { represented in Figure~\ref{fig:CMD_parabola_Levine} as green diamonds.  
Figure~\ref{fig:CMD_Levine}-right shows the data corrected for reddening, as in Figure~\ref{fig:CMD_parabola_Levine}. The Levine sources, shown here as blue dots, are clustered around the isochrone with dereddened magnitudes brighter than $m_{130}\simeq14$, corresponding to the H-burning limit on our 1~Myr isochrone. The two candidates M giants lie at the tip of the region containing galactic contaminants, consistent with the \cite{Levine+2006} classification.}

\begin{figure*}
\centering
\includegraphics[width=0.55\textwidth]{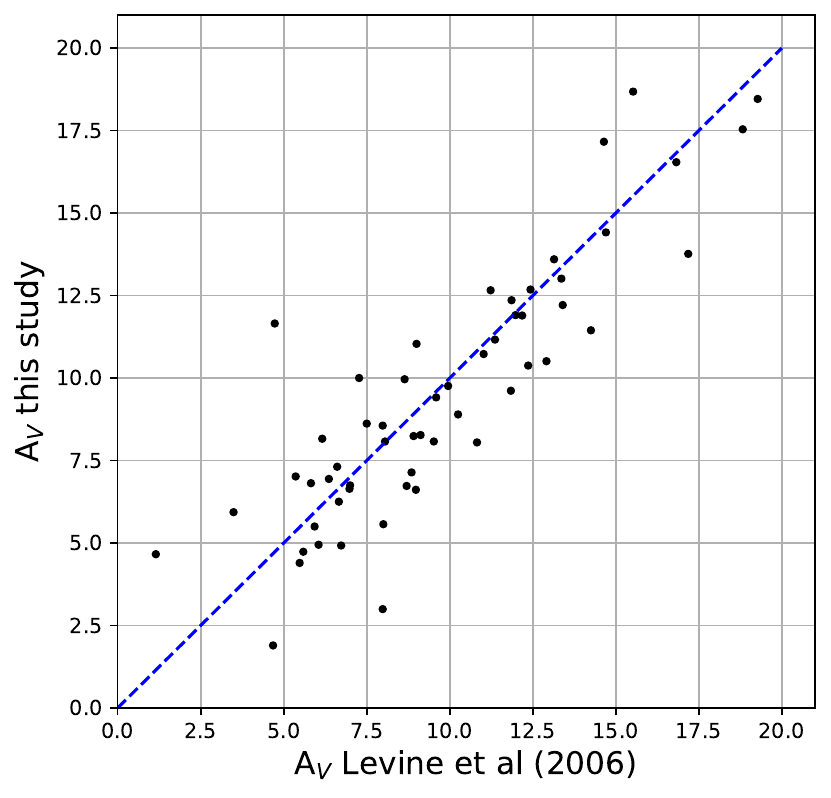}
\caption{Comparison between the $A_V$ values obtained using the WFC3 photometry in the $F139M$ filter, with our baseline reddening law vs. the values derived by \cite{Levine+2006} using ground-based near-IR spectroscopy.}
\label{fig:Av_comparison_vs_Levine}
\end{figure*}

Comparing  the stellar parameters estimated by \cite{Levine+2006} with those we determined in the previous sections, one has to take into account the different assumptions regarding reddening law, intrinsic stellar colors and families of evolutionary models. In Figure~\ref{fig:Av_comparison_vs_Levine}  we show the extinction values we derived using our { \cite{Cardelli_1989ApJ...345..245C} }reddening law vs. those of \cite{Levine+2006}. They determined $A_V$ by comparing their $J-H$ colors with those empirically determined by \cite{1992ApJS...82..351L, 1996ApJS..104..117L} and \cite{2002AJ....124.1170D}
and adopting the reddening law of \cite{1981ApJ...249..481C}, all to be consistent with their Flamingos photometric system. The plot shows strong correlation, no systematic differences and an average { scatter $A_V\simeq 2$ magnitudes vs. a mean value $A_V\simeq10$ magnitudes.} 

\begin{figure*}
\centering
\includegraphics[width=0.55\textwidth]{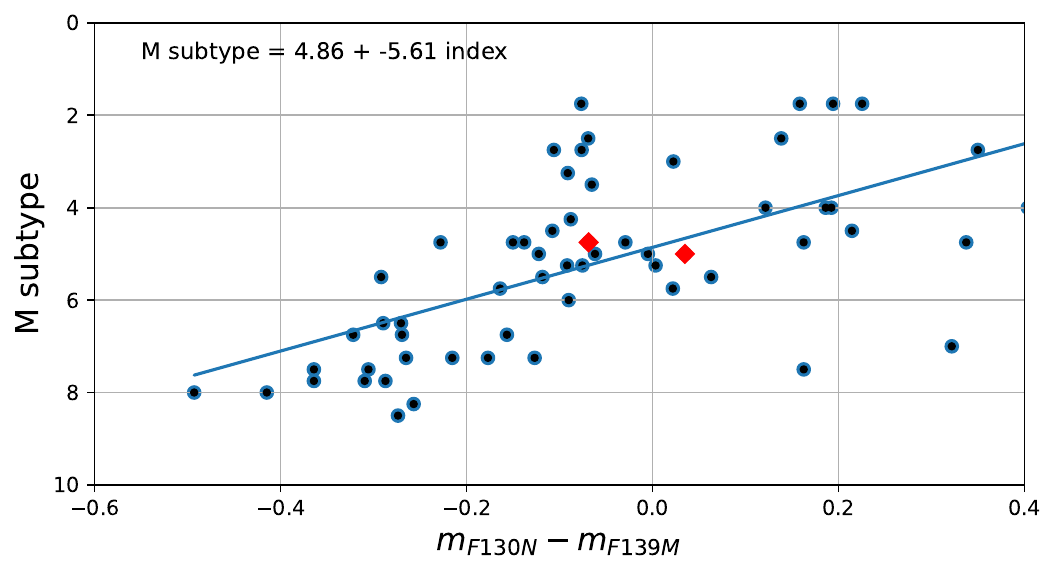}
\caption{Spectral classification reported by \cite{Levine+2006} 
 vs. the color index determined using our WFC3 photometry.}
\label{fig:Index_vs_subtype}
\end{figure*}

A more { direct} comparison is the one between the spectral types derived by \cite{Levine+2006} and our H$_2$O spectral index, presented in Figure~\ref{fig:Index_vs_subtype}. Here our index has been dereddened using again the { \cite{Cardelli_1989ApJ...345..245C}} reddening curve. The plot shows some significant scatter especially for positive values of our index, with the sources spectroscopically classified as early M-type ($<$M3) having index values ranging between -0.1 and 0.5. 
However, moving down and to the left in the plot, i.e. to lower mass and later spectral types { where the absorption feature is more prominent}, the correlation between spectral type and spectral index becomes stronger, with a standard deviation of about one sub-type. The linear fit plotted in Figure~\ref{fig:Index_vs_subtype}, with parameters shown on the plot, has been determined using all available data. 

\section{Conclusion}\label{sec:conclusion}
We have presented the results of a survey of the young ($\lesssim1$~Myr) stellar cluster NGC~2024 associated to the Flame Nebula in Orion. The data, taken with the Wide Field Camera 3 onboard of the Hubble Space Telescope, probe the $1.4~\mu$m H$_2$O absorption feature to discriminate the population of sub-stellar objects, down to a few \mjup, against highly reddened more massive stars.  

In a field of about 85 square arcmin we detect 808 point sources, 550 of them having sigma-to-noise $>5$ in all 3 filters. We estimate the reddening of the this smaller sample using a distance-independent two-color diagram, finding that the source distribution cannot be explained by trivially adopting a linear relation { between color and extinction} based on any of the standard reddening laws. Instead, given the high extinction, one has to account for the change of effective wavelength as the extinction increases. We therefore used synthetic photometry to show that the non-linearity of the reddening correction generally { matches the distribution of the sources in the diagram} and results in higher $A_V$ values as the reddening increases.  { Adopting two different reddening laws, the \cite{Fitzpatrick_1999PASP..111...63F} law with $R_V=2.5$ for a stellar photosphere at $T_{eff}=3000$~K and the \cite{Cardelli_1989ApJ...345..245C} law with $R_V=3.1$ for a stellar photosphere at $T_{eff}=10,000$~K, we compare  the resulting distributions of extinction values, generally  peaking at  $A_V\sim15$~mag}. The majority of highly reddened sources appear concentrated in the central part of field, dominated at visible wavelengths by an extended dark lane. { We then} reconstruct the dereddened color-magnitude diagrams and derive the luminosity histograms, plotting both the full sample of { candidate} cluster members and the extinction limited sub-sample containing 50\% of sources. We find that the { \cite{Fitzpatrick_1999PASP..111...63F} law, overestimating the extinction, produces an excess of luminous (therefore massive) stars not compatible with the standard Salpeter's slope of the IMF. The \cite{Cardelli_1989ApJ...345..245C} is in much better agreement with the canonical IMF slope, but still shows an excess of luminous stars in the full sample caused, which includes the highly reddened sources. The correlation between high extinction and luminosity may result from a residual underestimate of the extinction. On the other hand}, we posit that the correlation may real and due to the most embedded sources being younger and overluminous vs. the more evolved and less extinct { cluster stars in the foreground}. We compare our classification scheme based on the depth of the 1.4\um\, photometric feature with the results from the spectroscopic survey of \cite{Levine+2006}, finding general agreement especially for the late M subtypes, where the H$_2$O absorption feature is lmore prominent. Finally, we report a few peculiar sources and morphological features typical of the rich phenomenology commonly encountered in young star-forming regions.

\begin{acknowledgments}
{ The HST data presented in this paper were obtained from the Mikulski Archive for Space Telescopes (MAST) at the Space Telescope Science Institute. The specific observations analyzed can be accessed via \dataset[DOI: 10.17909/2ya9-db87]{https://doi.org/10.17909/2ya9-db87}.}
This work has made use of data from the European Space Agency (ESA)
mission {\it Gaia} 
(\url{https://www.cosmos.esa.int/gaia}), processed by
the {\it Gaia} Data Processing and Analysis Consortium (DPAC,
\url{https://www.cosmos.esa.int/web/gaia/dpac/consortium}). Funding
for the DPAC has been provided by national institutions, in particular
the institutions participating in the {\it Gaia} Multilateral Agreement.
{ We wish to acknowledge the anonymous referee for the helpful comments that helped improving the manuscript.}

\end{acknowledgments}

\appendix

\section{Individual sources and peculiar morphological features}\label{sec:individual}

Our choice of filters, aimed at discerning low-mass objects through the presence of H$_2$O molecular absorption and estimating their reddening, is less than ideal to trace the rich phenomenology generally associated with star-forming regions, better unveiled through narrow band imaging in recombination lines. Still, the visual inspection of the images reveals, especially in the western part of the nebula less affected by extinction, a number of features worthy of being reported. 

\begin{figure*}
\centering
\includegraphics[width=0.45\textwidth]{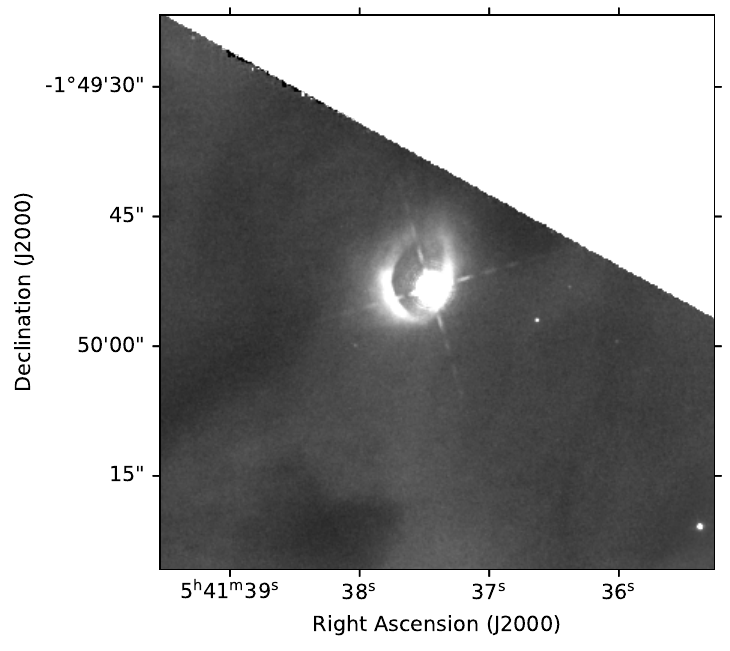}
\caption{Zoom-in on 2MASS J05413744-0149532 with its surrounding cavity.}
\label{fig:cavity}
\end{figure*}

Source IRS 1 appears encircled by an extended cavity about 3$^\prime$ or 1/3 of a parsec long, with edges traced by narrow, dark filaments (Figure~\ref{fig:cavity}). Proplyd 1 and 2 of \cite{Haworth+2021MNRAS.501.3502H} lie inside this cavity in the vicinity of IRS 1, consistent with their direct exposure to strong ultraviolet radiation; in our images they appear as point sources without any evidence of circumstellar material, proplyd 2 being coincidentally well aligned with a diffraction spike from IRS 1. The two other proplyds and the four candidate proplyds detected by \cite{Haworth+2021MNRAS.501.3502H} do not appear in our images.

\begin{figure*}
\centering
\includegraphics[width=0.55\textwidth]{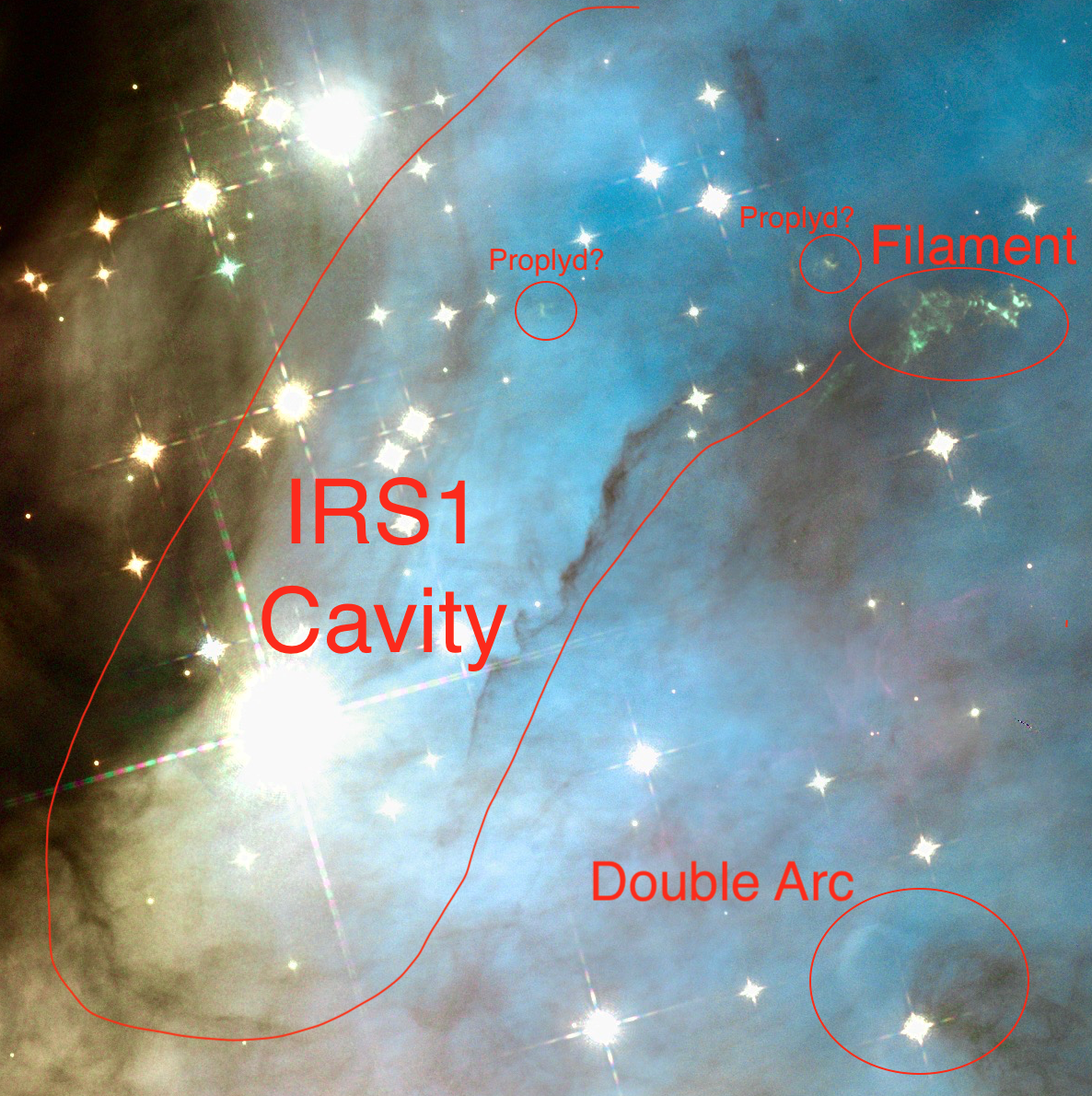}
\caption{Enlargement of the cavity surrounding the IRS 1 region, with the positions of the proplyds, filaments and arcs described in Figures~12, 13 and 14.}
\label{fig:IRS1_RGB}
\end{figure*}
\vfill\eject

\begin{figure*}
\centering
\includegraphics[width=0.75\textwidth]{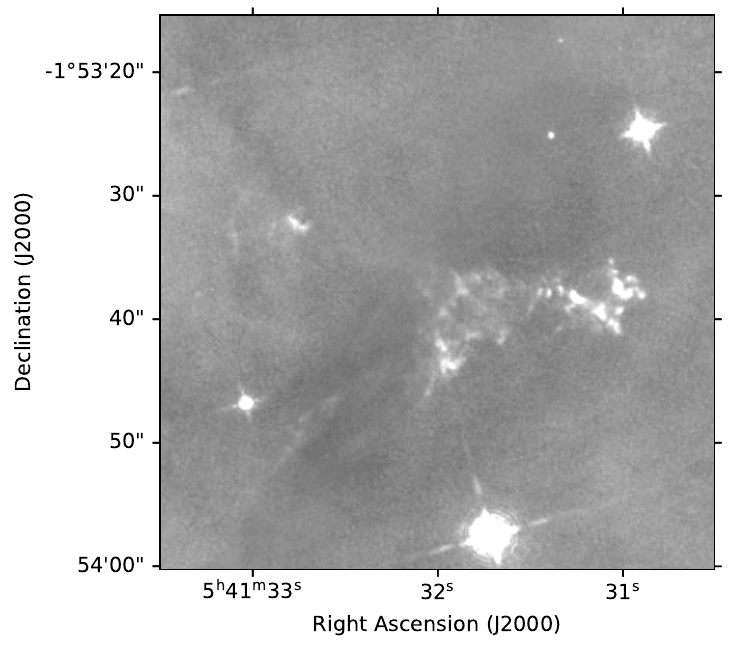}
\caption{Zoom-in onto the filamentary structure to the NW of the IRS 1, with the photoionized disk also visible in the upper-left quadrant.}
\label{fig:HH}
\end{figure*}

At the north-western end of the IRS 1 cavity, a complex filamentary structure (Figure~\ref{fig:HH}) approximately centered at RA=05:41:31.4 DEC=-01:53:38  (J2000.0) is most probably the working surface of a protostellar jet from an unidentified source. The filament is most prominent in the F130N filter. Nominally, this is a line-free filter, i.e., the Paschen-beta continuum, centered at 1,300~nm with a FWHM of 20~nm. The steep, blue cutoff of the filter at 1,290~nm should prevent the 1,282~nm Paschen-beta line from entering the passband, but our data suggest that some contamination may be present.

\begin{figure*}
\centering
\includegraphics[width=0.45\textwidth]{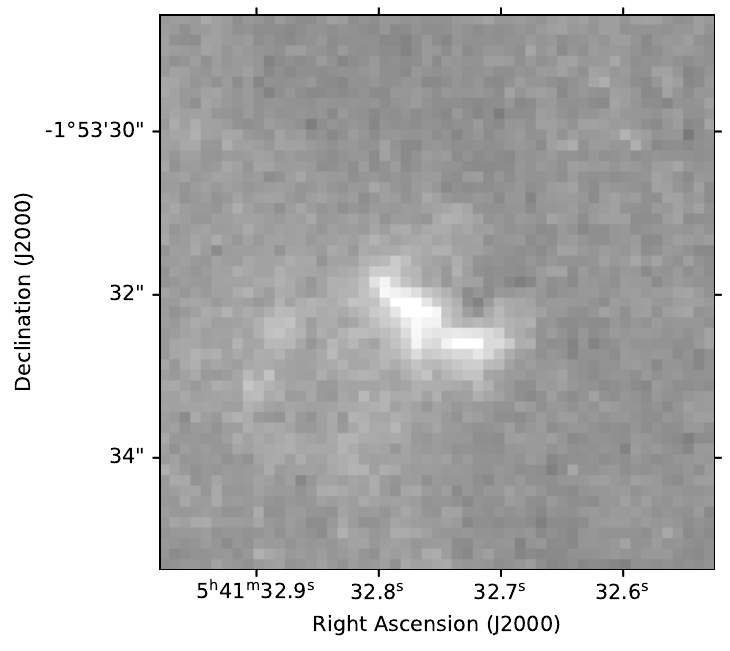}
\caption{Zoom-in on the binary proplyd at RA=05:41:32.7, DEC=-01:53:32. These images and the following ones are all in the F130N passband.}
\label{fig:BinaryProplyd}
\end{figure*}

\begin{figure*}
\centering
\includegraphics[width=0.45\textwidth]{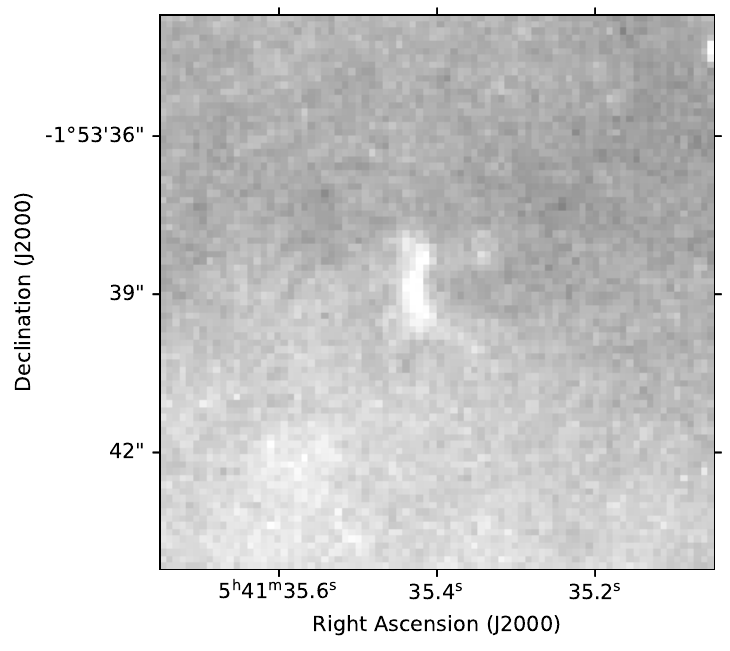}
\caption{Zoom-in on the proplyd within the IRS 1 cavity at RA=05:41:35.1, DWEC=-01:53:39. These images and the following ones are all in the F130N passband.}
\label{fig:brightrim_cavity}
\end{figure*}

Two other sources in the area show proplyd-like morphology. In particular at RA=05:41:32.75 DEC=-01:53:32.0 (J2000.0), about 10'' to the NE of the jet and within the field shown in Figure~\ref{fig:HH}, a compact extended source is characterized by a bright rim perpendicular to the direction of IRS 1. The enlargement shown in Figure~\ref{fig:BinaryProplyd} reveals that the rim may be broken in two parts, and one may recognize faint emission from the rear side opposite IRS 1, two compact jets protruding through the ionized rim and possibly two darker areas where one could expect finding dark disks, a morphology that would be consistent with a binary proplyd. A second compact arc of emission  (Figure~\ref{fig:brightrim_cavity}) also consistent with the presence of a proplyd, is visible within the boundaries of the cavity at RA=05:41:35.4 DEC=-01:53:39  (J2000.0). 

\begin{figure*}
\centering
\includegraphics[width=0.75\textwidth]{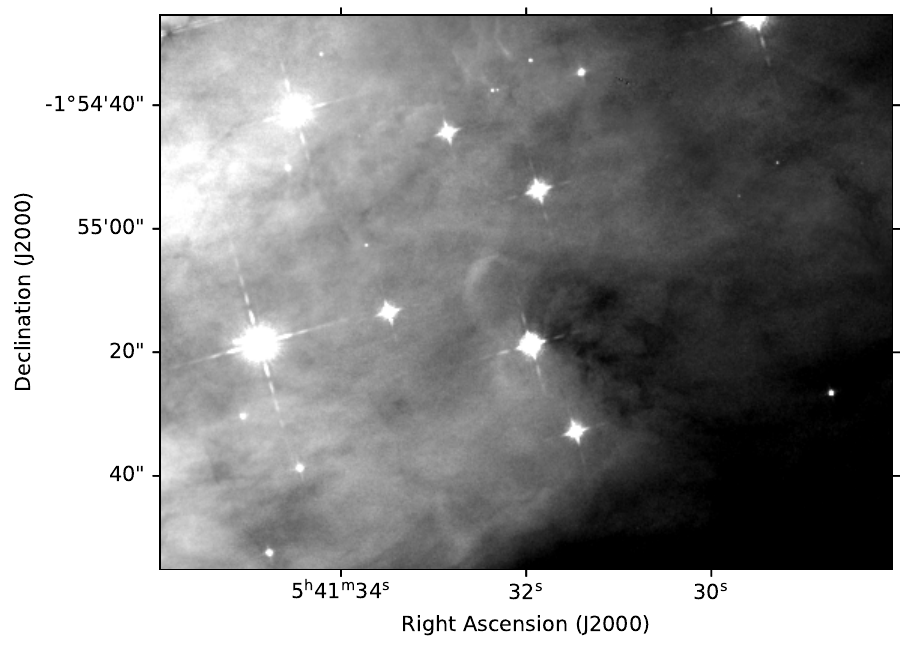}
\caption{Zoom-in the Double Arc to the West of IRS 1.}
\label{fig:Double Arc}
\end{figure*}

To the southwest of the cavity, a striking feature is a double pillar with a bright and dark component that appears aligned toward IRS 1 (Figure~\ref{fig:Double Arc}). A tempting explanation is that the bright arc traces the rim of a pillar illuminated by IRS 1, a pillar which casts its shadow on a translucent veil behind it that therefore traces in silhouette the morphology of the underlying illuminated structure, providing us with a rare direct glimpse of the 3D structure of the region.

At the northern edge of our survey field, our source \#810 (2MASS J05413744-0149532) at RA=
05:41:37.4 DEC=-01:49:53.7 (J2000.0) is encircled by a 6'' $\times$ 9'' ellipsoidal  cavity with sharp inner boundaries and no indication of bipolar morphology (Figure~\ref{fig:cavity}).

\begin{figure*}
\centering
\includegraphics[width=0.45\textwidth]{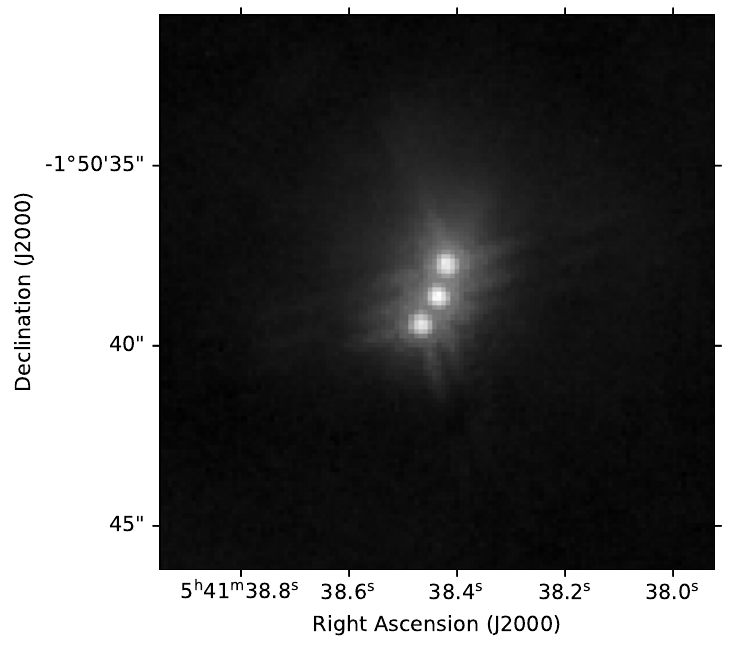}
\caption{Zoom-in on a remarkable triple system.}
\label{fig:3stars}
\end{figure*}

Also in the vicinity of the northern edge of our survey field, a source at RA=05:41:38.44 DEC= -01:50:38.5 (J2000.0) is resolved into a peculiar triple system of nearly co-aligned and almost equal brightness ($m_{130}\simeq 13.0$~mag) stars (Figure~\ref{fig:3stars}). These sources (\#776, 778 and 780 in our catalog) have $\simeq$0.8'' separation between each pair, corresponding to a projected distance of about 320~A.U.

\vfill\eject
\FloatBarrier
\bibliography{sample63}
\bibliographystyle{aasjournal}



\end{document}